\documentclass[sigconf,natbib=true]{acmart}
\usepackage{soul}
\AtBeginDocument{%
  \providecommand\BibTeX{{%
    \normalfont B\kern-0.5em{\scshape i\kern-0.25em b}\kern-0.8em\TeX}}}

\newcommand\oconcat{\ensuremath{\mathchoice%
    {\rotatebox[origin=c]{90}{$\ominus$}}%
    {\rotatebox[origin=c]{90}{$\ominus$}}%
    {\rotatebox[origin=c]{90}{$\scriptstyle\ominus$}}%
    {\rotatebox[origin=c]{90}{$\scriptscriptstyle\ominus$}}%
}}

\setcopyright{acmcopyright}
\copyrightyear{2023}
\acmYear{2023}
\setcopyright{acmlicensed}
\acmConference[CIKM '23]{Proceedings of the 32nd ACM International Conference on Information and Knowledge Management}{October 21--25, 2023}{Birmingham, United Kingdom}
\acmBooktitle{Proceedings of the 32nd ACM International Conference on Information and Knowledge Management (CIKM '23), October 21--25, 2023, Birmingham, United Kingdom}\acmDOI{10.1145/3583780.3615512}
\acmISBN{979-8-4007-0124-5/23/10}
\acmPrice{15.00}

\begin{document}

\title{CPMR: Context-Aware Incremental Sequential Recommendation with Pseudo-Multi-Task Learning}

\author{Qingtian Bian}
\affiliation{%
  \institution{Nanyang Technological University}
  \country{Singapore}}
\email{bian0027@e.ntu.edu.sg}

\author{Jiaxing Xu}
\affiliation{%
  \institution{Nanyang Technological University}
  \country{Singapore}}
\email{jiaxing003@e.ntu.edu.sg}

\author{Hui Fang}
\affiliation{%
  \institution{Shanghai University of Finance and Economics}
  \country{China}}
\email{fang.hui@mail.shufe.edu.cn}

\author{Yiping Ke}
\affiliation{%
  \institution{Nanyang Technological University}
  \country{Singapore}}
\email{ypke@ntu.edu.sg}

\begin{abstract}
The motivations of users to make interactions can be divided into static preference and dynamic interest. To accurately model user representations over time, recent studies in sequential recommendation utilize information propagation and evolution to mine from batches of arriving interactions. However, they ignore the fact that people are easily influenced by the recent actions of other users in the contextual scenario, and applying evolution across all historical interactions dilutes the importance of recent ones, thus failing to model the evolution of dynamic interest accurately. To address this issue, we propose a \textbf{C}ontext-Aware \textbf{P}seudo-\textbf{M}ulti-Task \textbf{R}ecommender System (CPMR) to model the evolution in both historical and contextual scenarios by creating three representations for each user and item under different dynamics: static embedding, historical temporal states, and contextual temporal states. To dually improve the performance of temporal states evolution and incremental recommendation, we design a Pseudo-Multi-Task Learning (PMTL) paradigm by stacking the incremental single-target recommendations into one multi-target task for joint optimization. Within the PMTL paradigm, CPMR employs a shared-bottom network to conduct the evolution of temporal states across historical and contextual scenarios, as well as the fusion of them at the user-item level. In addition, CPMR incorporates one real tower for incremental predictions, and two pseudo towers dedicated to updating the respective temporal states based on new batches of interactions. Experimental results on four benchmark recommendation datasets show that CPMR consistently outperforms state-of-the-art baselines and achieves significant gains on three of them. The code is available at: \url{https://github.com/DiMarzioBian/CPMR}.
\end{abstract}



\maketitle

\section{Introduction}
For human beings, making interactions is an important behavior to understand other objects and update their own recognitions. With the time information available, interaction trajectories are typically modeled as chronologically ordered sequences in real applications, e.g., e-commerce (click, add to cart, buy, and even neglect while browsing) \cite{zhou2018deep}, music apps (listen for a while or switch rapidly) \cite{brost2019music}, as well as social media (post, reply, forward, and mention) \cite{de2013anatomy}. The motivations for making an interaction vary but can be generally categorized into two types: static preference (long-term interest) and dynamic interest (short-term interest). How to properly model these two types of user interest from interaction sequences breeds the need for effective sequential recommender models.

Different from traditional sequential recommender systems \cite{tang2018personalized, kang2018self, ma2020memory, chang2021sequential} that only make use of the chronological order of interactions, recent studies \cite{li2020time, fan2021continuous} have proven that timestamps of interactions are more informative in characterizing temporal dynamics. In terms of time modeling, some representation learning works employ additive or concatenative temporal embeddings \cite{xu2019self, li2020time, fan2021continuous, yang2022stam}. Some further model the continuous decay in the effect of interactions after their occurrences \cite{cho2020meantime, zhou2021temporal}. Moreover, to better capture the sequential information from discrete occurrences of interactions in continuous time, hybrid models mixing temporal point processes and recommender systems \cite{dai2016deep, vassoy2019time, bai2019ctrec, wang2021sequential} manage to calculate time-based intensity on items to give recommendations. Some graph-based works \cite{wang2020next, fan2021continuous} also introduce time-based graphs to model complex dynamic connectivities and adopt additive or concatenative temporal embeddings. The above-mentioned studies model temporal representations in the same way as static embeddings, and thus fail to capture complex dynamics in time-varying interests.

To fully represent the dynamic states embeddings over time, IncCTR \cite{wang2020practical} introduces incremental learning in Click-Through Rate prediction. SML \cite{zhang2020retrain} and FIRE \cite{xia2022fire} consider the time efficiency and train their models in a fast incremental learning manner without querying historical data. Besides these works, two evolution models for incremental sequential recommendation, JODIE \cite{kumar2019predicting} and CoPE \cite{zhang2021cope}, have been proposed. JODIE employs a recurrent neural network structure to discretely model the interest trajectories on dynamic embeddings. CoPE approximates the continuous evolution between sets of concurrent interactions by using CGNN \cite{fan2021continuous} to aggregate all historical interactions. Both evolution models learn static embeddings and temporal states, representing their static attributes and dynamic properties as functions of time, respectively.

Despite their superiority over models without considering temporal states, these evolution models evolve their temporal states across the entire history graph without specially treating recent interactions. Consequently, they hardly capture the changing dynamics and ignore the timeliness of temporal states, that is, the impact of recent contexts on users' dynamic interests. Considering the fact that such contexts are sensitive to time, we propose a new evolution model, namely \textbf{C}ontext-Aware \textbf{P}seudo-\textbf{M}ulti-Task \textbf{R}ecommender System (CPMR), for fusing the evolution of both historical and contextual scenarios in an MTL-like manner. Specifically, CPMR creates three representations for each user and item under different dynamics: static embedding, historical temporal states, and contextual temporal states. Based on them, CPMR carries out effective information fusion and efficient joint optimization by designing a Pseudo-Multi-Task-Learning (PMTL) paradigm.

\begin{figure}[t]
\centering
\vspace{-0.1cm}
\setlength{\abovecaptionskip}{-0.0cm}
\includegraphics[width=0.6\linewidth]{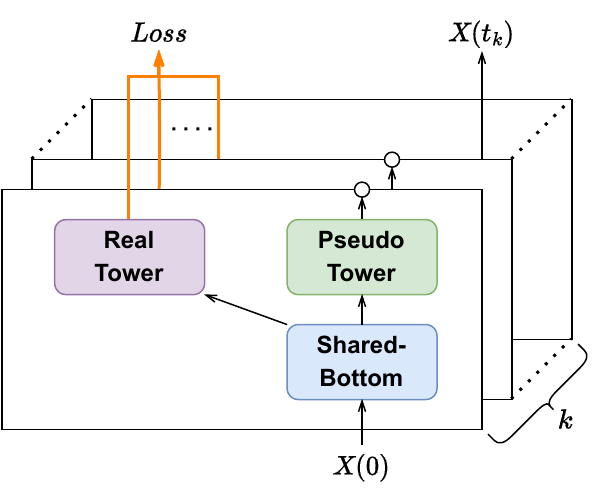}
\caption{A simple illustration of our proposed Pseudo-Multi-Task Learning paradigm over incremental recommendation task.}
\label{fig:pmtl}
\vspace{-0.2cm}
\end{figure}

MTL is a natural and principled solution to our recommendation problem as our modeling involves multiple tasks for evolving, updating, and fusing different temporal dynamics, all sharing the same set of temporal states. Conventional MTL cannot be directly applied because not all of our tasks generate losses. Therefore, we devise a new pseudo-MTL specifically for our problem. As shown in Figure \ref{fig:pmtl}, tasks of our proposed PMTL can be divided into the real task that generates losses and pseudo tasks that generate updated input for the next recurrence.

CPMR is an evolution model implemented under this PMTL paradigm by recurrently evolving the temporal state. In particular, CPMR leverages recommendation as a real task within the prediction module of the real tower, while employing the updating of temporal states as pseudo tasks within the update modules of the simulated towers. In order to model the continuous evolution and the mutual information sharing of historical and contextual temporal states, CPMR also employs two evolution module instantiations and two fusion module instantiations as the shared-bottom networks in PMTL paradigm. More specifically, the update module captures the instant evolution in temporal states from concurrent interactions, while the evolution module focuses on modeling the continuous evolution in temporal states from intervals between batches of interactions. To distinguish evolution within historical and contextual scenarios, CPMR creates two instances of both modules, each dedicated to leveraging the history graph and the context graph, respectively. In particular, the history graph consists of all the interactions that have occurred, and the context graph consists of the interactions that occurred within a fixed-length sliding time window (context window) from the current moment. After evolution, these temporal states are mutually updated at the user-item levels in two fusion module instantiations.

Empowered by the PMTL paradigm and context awareness, CPMR is able to dynamically model the historical and contextual temporal states of users and items and make recommendations. Experimental results demonstrate that CPMR outperforms state-of-the-art baselines and achieves 30.98\% gains on MRR and 27.39\% gains on Recall@10. Our contributions are summarized as follows:
\begin{itemize}
\item We propose a Pseudo-Multi-Task-Learning module to stack single-target incremental recommendations into one multi-target task by mutually evolving temporal states between each task.
\item We devise CPMR based on the PMTL paradigm, which enables the evolution and fusion of user interests and item attributes as temporal states within both historical and contextual scenarios.
\item We conduct extensive experiments on four datasets to evaluate the effectiveness of CPMR and perform ablation studies on fusion modules and proposed contextual temporal states. The results show that CPMR consistently outperforms state-of-the-art baselines, where both proposed components play important roles.
\end{itemize}

\section{Related Work}
\textbf{Sequential Recommendation (SR)}: it is a task to predict the next behavior leveraging from the sequence of chronologically ordered historical behaviors. The earliest work, FPMC \cite{rendle2010factorizing}, utilized the Markov Chain to model the transition pattern within sequences. To harness the representation learning capability of deep learning, CNN-based CASER \cite{tang2018personalized} viewed sequential embeddings as images to extract information. Recent sequential recommender works can be divided into two categories: recurrent-based methods \cite{wu2017recurrent, zhu2017next, xu2019recurrent, donkers2017sequential, zheng2022disentangling} and attention-based methods \cite{ying2018sequential, kang2018self, sun2019bert4rec, zhang2019feature, zhou2018deep, zhou2019deep}.

Some sequential recommender systems explore temporal information to enhance representation learning. For example, \cite{xu2019self} conducted extensive experiments on different types of temporal embeddings. TiSASRec \cite{li2020time} embedded relative time intervals to associate with time. CTRec \cite{bai2019ctrec} and DeepCoevolve \cite{dai2016deep} employed temporal point processes to introduce the temporal dynamics into recommendations. JODIE \cite{kumar2019predicting} represented temporal states of users and items by embedding trajectories. However, the update mechanisms for temporal states in these models largely rely on the entire history while ignoring the influence of context. To address this limitation, we calibrate the evolution by exploiting both historical and contextual information in a more timely manner. 

\addtolength{\parskip}{4pt}
\noindent\textbf{Graph-based Recommendation}: As each sequence can be viewed as a subgraph, the recommendation dataset can be transformed into a user-item bipartite graph or an item-item graph. For instance, SR-GNN \cite{wu2019session} firstly introduced GNN techniques into recommendation tasks. LightGCN \cite{he2020lightgcn} designed a simple but effective graph convolution network (GCN). Subsequently, a number of studies apply graph learning to a variety of different recommendation tasks \cite{song2019session, zhou2020improving, wu2019session, ma2020memory}, showing the potential of this combination.

Besides, temporal information can also be applied to graph-based recommendations. TGSRec \cite{fan2021continuous} unified sequential patterns and temporal collaborative signals to improve recommendation. CoPE \cite{zhang2021cope} proposed a CGNN-based method to learn from continuous propagation and evolution. FIRE \cite{xia2022fire} designed graph filters from a graph signal processing perspective to capture the temporal dynamics and address the cold-start problem. RETE \cite{wang2022rete} proposed a retrieval-enhanced recommendation model based on knowledge graphs to model the temporal dynamics. However, the number of new interactions at each timestamp is too small compared to all historical interactions. Hence applying GNN directly on the graph of all historical interactions cannot effectively capture the dynamics of users and items. On the contrary, in this work, we propose to mitigate this problem by applying another GNN on a more dynamic context graph.

\addtolength{\parskip}{4pt}
\noindent\textbf{Multi-Task Recommendation}: Multi-Task Learning (MTL) is an active research topic in recommender systems, drawing attention from both industry and research. The general network architecture of MTL \cite{ma2018entire, ma2018modeling, tang2020progressive, sun2020learning, bai2022contrastive} consists of a shared bottom network that learns task-shared knowledge and multiple task-specific towers to generate the results required by respective tasks. According to the setting of the targets, MTL models can be divided into two categories, one is to use auxiliary tasks to assist in optimizing single or multiple target tasks \cite{wang2019multi, wu2019feature}, and the other is to optimize all tasks at the same time \cite{gao2019neural, zhu2019dtcdr}. By definition and task settings, common single-task problems do not have to be transformed into multi-task problems. However, if we consider the evolving temporal states as inputs to the shared-bottom network, one tower making recommendations and the other towers updating the input temporal states for further recommendations, we can jointly optimize these single-task recommendations in an MTL-like way by sharing the evolving temporal states among them.

\section{Preliminaries}
\subsection{Problem Formulation}
We provide a formal definition of the incremental sequential recommendation (ISR) task we are tackling. Assume we have a set of users $\mathcal{U}$ and a set of items $\mathcal{I}$. We see each user-item interaction with a timestamp as a triplet $(u, i, t)$. Therefore, at a given timestamp $t_k$, a user $u_p \in \mathcal{U}$ has a chronologically ordered historical interaction sequence $S_p=\{(u_p, i_1, t_1), ..., (u_p, i_{k-1}, t_{k-1})\}$ and $0 = t_0 \leq t_1 \leq t_2 \leq ... \leq t_{k-1}$. If $u_p$ makes an interaction at $t_k$, ISR aims to predict the ground-truth item $i_k$ that $u_p$ will interact with at $t_k$ by mining $u_p$'s interest from $S_p$ together with all observed interactions from other users before $t_k$. That is to say, in terms of time, this task strictly adheres to the principle of no data leakage. Unlike some other incremental recommendation tasks \cite{wang2020practical, xia2022fire}, the ISR task allows all history data to be used, rather than using only the model's current states and incoming interactions.  The problem formulation of incremental sequential recommendation is given as:

\noindent
\textbf{Input:} The historical interactions of all users before timestamp $t_k$. \\
\textbf{Output:} A recommender system that estimates the probability of user $u_p$ interacting with every candidate item $i\in\mathcal{I}$ at $t_k$, and recommends a top $N$ recommendation list with the highest probabilities to user $u_p$.

Based on the definition of ISR, JODIE, CoPE and our CPMR can all be classified as ISR models.

\begin{figure}[t]
\centering
\setlength{\abovecaptionskip}{0.0cm}
\includegraphics[width=\linewidth]{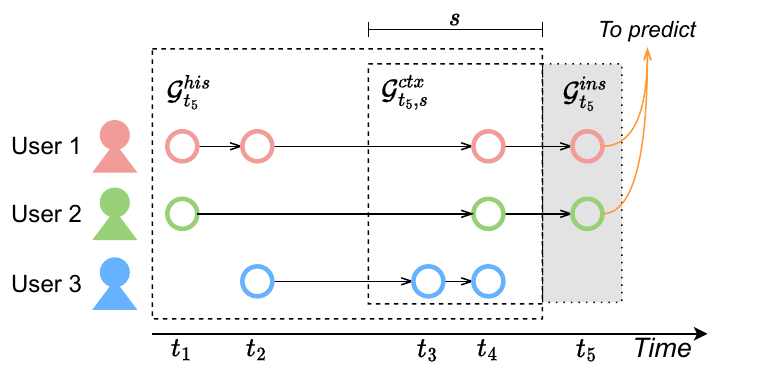}
\caption{Illustration of the history graph, context graph and instant graph.}
\label{fig:timeline}
\end{figure}

\begin{figure*}[h]
\centering
\includegraphics[width=\linewidth]{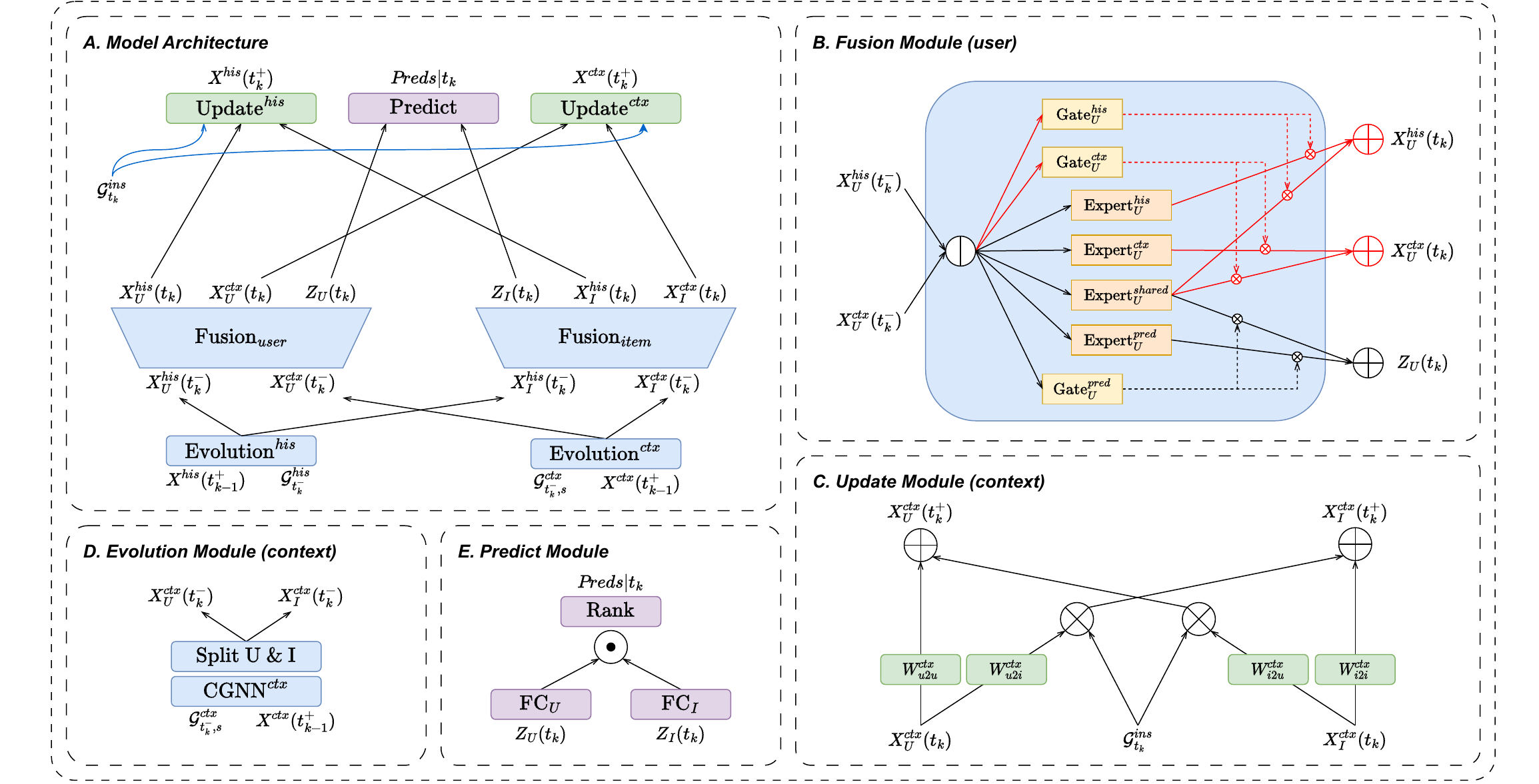}
\caption{The proposed structure of Context-Aware Pseudo-Multi-Task Recommender System. In subfig A, The shared-bottom network consists of all blue blocks, and each green or pink block represents a pseudo tower or a real tower.}
\label{fig:model}
\end{figure*}
\subsection{Graph Formulation}
By joining the interactions of all sequences into one edge set $\mathcal{E}= S_1 \cup S_2 \cup ... \cup S_n$,  we can represent this edge set as a bipartite interaction graph $\mathcal{G} = (\mathcal{U} \cup \mathcal{I}, \mathcal{E})$, in which $\forall u \in \mathcal{U}$, all its neighbor nodes $\mathcal{N}(u)$ are of item type, i.e., $\mathcal{N}(u) \subseteq \mathcal{I}$, and vice versa. Without loss of generality, we normalize the timestamps in the dataset into $[0, 1]$ and define the moments right before and right after the timestamp $t_k$ as $t_k^-$ and $t_k^+$. As shown in Figure \ref{fig:timeline}, we further define three types of graphs based on the time spans in CPMR.

\begin{definition}[Instant Graph]
    Given an interaction graph $\mathcal{G} = \left(\mathcal{U} \cup \mathcal{I}, \mathcal{E}\right)$, the instant graph at timestamp $t_k$ is formed by all the interactions happened right at $t_k$, i.e., $\mathcal{G}^{ins}_{t_k} = \left(\mathcal{U} \cup \mathcal{I}, \mathcal{E}^{ins}_{t_k}\right)$ and $\mathcal{E}^{ins}_{t_k} = \left\{(u', i', t')\in \mathcal{E} | t' = t_k\right\}$.
\end{definition}

\begin{definition}[History Graph]
    Given an interaction graph $\mathcal{G} = \left(\mathcal{U} \cup \mathcal{I}, \mathcal{E}\right)$, the history graph at timestamp $t$ is formed by all the interactions happened before $t$, i.e., $\mathcal{G}^{his}_{t} = \left(\mathcal{U} \cup \mathcal{I}, \mathcal{E}^{his}_{t}\right)$ and $\mathcal{E}^{his}_{t} = \left\{(u', i', t')\in \mathcal{E} | t' < t\right\}$. Each history graph stays unchanged between two adjacent interactions, i.e., for $\forall t \in (t_{k-1}, t_k)$, $\mathcal{G}^{his}_{t} = \mathcal{G}^{his}_{t_{k-1}^+} = \mathcal{G}^{his}_{t_k^-}$.
\end{definition}

\begin{definition}[Context Graph]
    Given an interaction graph $\mathcal{G} = \left(\mathcal{U} \cup \mathcal{I}, \mathcal{E}\right)$ and the length of context window $s$, the context graph at timestamp $t$ is formed by all the interactions happened between the interval $[t-s, t)$, i.e., $\mathcal{G}^{ctx}_{t, s} = \left(\mathcal{U} \cup \mathcal{I}, \mathcal{E}^{ctx}_{t, s}\right)$ and $\mathcal{E}^{ctx}_{t, s} = \left\{(u', i', t')\in \mathcal{E} | t - s \leq t' < t\right\}$, and $\mathcal{E}^{ctx}_{t, s} \subseteq \mathcal{E}^{his}_t$.
\end{definition}

Given an interaction graph $\mathcal{G} = \left(\mathcal{U} \cup \mathcal{I}, \mathcal{E}\right)$, its adjacency matrix is denoted by $\boldsymbol{adj} = \left[\begin{matrix}0 & \boldsymbol{B}\\ \boldsymbol{B}^T & 0 \end{matrix}\right]$, where $\boldsymbol{B}\in \mathbb{R}^{|\mathcal{U}| \times |\mathcal{I}|}$ is the bi-adjacency matrix containing all user-item interactions in the interaction graph:

\begin{equation}
\boldsymbol{B}_{u, i} = \left\{
\begin{aligned}
&1\quad \text{if}\ (u, i, t) \in \mathcal{E}, \\
&0\quad \text{otherwise}.
\end{aligned}
\right.
\end{equation}
Because of the sparsity problem in the recommendation dataset, the degrees of nodes can be very different. Therefore we normalize the adjacency matrix and give it as follows:
\begin{equation}
    \boldsymbol{\widetilde{adj}} := \frac{\alpha_0}{2}(\boldsymbol{I} + \boldsymbol{D}^{-\frac{1}{2}}\cdot \boldsymbol{adj} \cdot \boldsymbol{D}^{-\frac{1}{2}}),
\label{eq:adj}
\end{equation}
where $\boldsymbol{D}$ denotes the degree matrix of $\boldsymbol{adj}$, and $\alpha_0$ is set to 0.98 to make sure all eigenvalues of $\widetilde{\boldsymbol{adj}}$ to be in the interval $[0, 1)$ for future modeling and approximations \cite{chung1997spectral}.

\section{Methodology}
\subsection{Pseudo-Multi-Task Learning Paradigm}
Previous evolution models, JODIE \cite{kumar2019predicting} and CoPE \cite{zhang2021cope}, recurrently evolve temporal states to make recommendations as they only model a single type of temporal dynamics,  i.e., historical dynamics. In this work, we propose to model contextual dynamics in addition to the historical one. By doing so, three different tasks naturally arise: instant update of contextual/historical temporal dynamics, the fusion and continuous evolution of these two types of temporal dynamics, as well as the recommendation task. Beneath all the tasks, they share the same set of user/item temporal states. Meanwhile, each individual task has its own characteristics and target. This motivates us to employ Multi-Task Learning (MTL) as a principled solution. However, each task in conventional MTL generates its own loss, while this is not the case in our problem. Therefore, we design a new pseudo-MTL (PMTL) paradigm to better fit our need in ISR. Specifically, we designate a task that generates losses as a real task, while those not bound with losses as pseudo tasks. Figure \ref{fig:pmtl} illustrates our proposed PMTL paradigm over the task of ISR. The shared-bottom network handles the fusion and evolution of temporal states. The output of shared-bottom network are feed into the pseudo tasks for instant updates and to the real task for prediction. In this way, we achieve a joint optimization akin to MTL.

In this section, we devise CPMR based on the PMTL paradigm, a system modeling both contextual and historical temporal states to solve the ISR task. As shown in Figure \ref{fig:model}.A, CPMR consists of four different types of modules, each tackling different tasks. (1) Evolution Module: to perform continuous update of temporal states in intervals between batches of interactions. (2) Fusion Module: to fuse information from different dynamics scenarios at the user-item level. (3) Update Module: to conduct instant update of temporal states based on new batches of interactions. (4) Predict Module: to make incremental recommendations. Following our PMTL paradigm, CPMR instantiates two evolution modules and two fusion modules as the shared-bottom network (blue blocks in Figure \ref{fig:model}.A). With the temporal states generated by the shared-bottom network, CPMR further implements two instantiations of the update module as two pseudo tasks (green blocks) to perform instant updates on temporal states. Finally, CPMR instantiates the real task of PMTL by a predict module (pink block), which generates recommendation losses. While the pseudo tasks do not directly generate losses, its updated temporal states make the next inputs different from the current inputs, and thus the next recurrence's loss will be different. By jointly optimizing these losses from different recurrences, our model can not only be efficiently trained with less time of backpropagation but also reduce overfitting on specific recurrences.

\subsection{Overview of CPMR}
\begin{figure}[t]
\centering
\setlength{\abovecaptionskip}{-0.0cm}
\includegraphics[width=\linewidth]{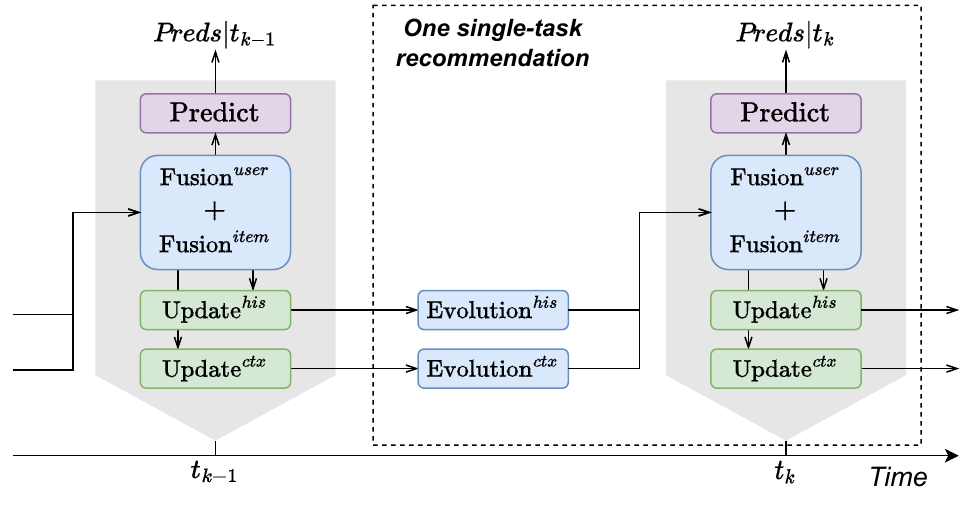}
\caption{Workflow of the CPMR. Modules in the same grey block are executed in a short time.}
\label{fig:workflow}
\vspace{-0.2cm}
\end{figure}

To be specific, CPMR learns three sets of vector representations for each user and item, static embeddings, historical temporal states and contextual temporal states, which capture the basic preference (attributes) and the time-varying dynamics in historical and contextual scenarios respectively. On top of these representations we implement the aforementioned modules as:
\begin{itemize}
    \item{\textbf{Evolution Module}} Through two parallel GNN encoders, the historical and contextual temporal states of users and items evolve from $t^+_{k-1}$ to $t^-_{k}$. This module contains two instantiations respectively for historical and contextual scenarios, each incorporating the corresponding input graph. The module is designed to approximate the continuous states evolution at interval $(t^+_{k-1}, t^-_k)$.
    \item{\textbf{Fusion Module}} The mutual update of temporal states is performed in this module. It takes the latest historical and contextual temporal states at $t^-_k$ as inputs and generates final users' and items' representations for recommendations at $t_k$, and the updated historical and contextual temporal states at $t_k$ via information fusion. Unlike in the evolution module, the two instantiations in this module are tailored for users and items respectively. They aim to blend the information from both historical and contextual sources and inject such information into the temporal states of each type of nodes, i.e., either user or item nodes.
    \item{\textbf{Predict Module}} The top-N recommendation list is generated by the module. It takes the final users' and items' representations generated by the fusion module at $t_k$ as inputs.
    \item{\textbf{Update Module}} The update module represents newly arrived concurrent interactions at $t_k$ as an instant graph and discretely updates the relevant temporal states from $t_k$ to $t^+_k$. Same as in the evolution module, this module has two instantiations respectively for historical and contextual dynamics.
\end{itemize}
As shown in Figure \ref{fig:workflow}, CPMR runs the evolution module, the fusion module, and the update module in a loop to keep the temporal states updated. According to the model structure illustrated in Figure \ref{fig:model}, for $k=1,2,...,n$, we can write CPMR in a recurrent form, ignoring the notations of users and items:
\begin{equation}
\left\{
\begin{aligned}
&\boldsymbol{X}^{his}(0^+) = \boldsymbol{X}^{ctx}(0^+) = \boldsymbol{E} \\
&\boldsymbol{X}^{his}(t_k^-) = \text{Evolution}^{his}\left(\mathcal{G}^{his}_{t_{k}^-},\ \boldsymbol{X}^{his}(t_{k-1}^+),\ t_{k-1},\ t_{k} \right) \\
&\boldsymbol{X}^{ctx}(t_k^-) = \text{Evolution}^{ctx}\left(\mathcal{G}^{ctx}_{t_{k}^-, s},\ \boldsymbol{X}^{ctx}(t_{k-1}^+),\ t_{k-1},\ t_{k} \right) \\
&\boldsymbol{X}^{his}(t_k),\ \boldsymbol{X}^{ctx}(t_k),\ \boldsymbol{Z}(t_k) = \text{Fusion}\left(\boldsymbol{X}^{his}(t_k^-),\ \boldsymbol{X}^{ctx}(t_k^-)\right) \\
&\boldsymbol{X}^{his}(t_k^+) = \text{Update}^{his}\left(\mathcal{G}^{ins}_{t_{k}},\ \boldsymbol{X}^{his}(t_k)\right) \\
&\boldsymbol{X}^{ctx}(t_k^+) = \text{Update}^{ctx}\left(\mathcal{G}^{ins}_{t_{k}},\ \boldsymbol{X}^{ctx}(t_k)\right)
\end{aligned}
\right.
\label{eq:sys_form}
\end{equation}
where $\boldsymbol{E} \in \mathbb{R}^{|\mathcal{U} \cup \mathcal{I}|\times d}$ denotes the static embeddings for all users and items, $\boldsymbol{X} \in \mathbb{R}^{|\mathcal{U} \cup \mathcal{I}|\times d}$ denotes the users' and items' temporal states, $\boldsymbol{Z} \in \mathbb{R}^{|\mathcal{U} \cup \mathcal{I}|\times d}$ denotes the users' and items' final representations for making recommendations, and the Fusion function here denotes two fusion module instantiations of user or items for simplicity. In our implementation, each expert and gating network is a linear layer.

One example of the execution flow at the recurrence of $t_k$ is shown in Figure \ref{fig:workflow}, and summarized as follows. 
\begin{enumerate}
    \item Right after some interactions were made at $t_{k-1}$, CPMR runs the evolution module to update temporal states from $\boldsymbol{X}(t_{k-1}^+)$ to $\boldsymbol{X}(t_{k}^-)$ at an interval in-between adjacent interactions $(t_{k-1}^+, t_{k}^-)$.
    \item Right before some interactions are made at $t_k$, CPMR first calls the fusion module to update evolved temporal states from $\boldsymbol{X}(t_{k}^-)$ to $\boldsymbol{X}(t_{k})$ via information fusion and then calls the predict module to make top-N recommendations for each user involved in the coming interactions $\mathcal{E}^{ins}_{t_k}$.
    \item At timestamp $t_k$, these interactions are made and the instant graph $\mathcal{G}^{ins}_{t_k} = \left(\mathcal{U} \cup \mathcal{I}, \mathcal{E}^{ins}_{t_k}\right)$ is constructed. If CPMR accumulates losses for a given number of recurrences based on the predictions in Step (2) and the instant graph $\mathcal{G}^{ins}_{t_k}$, it will perform TBPTT to update the learnable parameters. Besides that, CPMR also runs the update module to add the effects of this set of interactions $\mathcal{E}^{ins}_{t_k}$ into temporal state $\boldsymbol{X}(t_k)$ to get $\boldsymbol{X}(t_k^+)$. After this, CPMR moves to the next recurrence.
\end{enumerate}
We will introduce each module in detail in subsequent subsections.

\subsection{Evolution Module}
In the evolution module, the goal is to model new temporal states $\boldsymbol{X}^{his}(t)$ and $\boldsymbol{X}^{ctx}(t)$ for $t \in (t_{k-1}^+, t_k^-)$. Essentially, it models the interest propagation that exists in an interval between adjacent interactions (i.e., an interval without any interactions). Instead of conducting aggregation on all happened interactions ignoring time gaps as what CoPE does, we propose mining from trendy items and their corresponding users by establishing a context environment within the context window and removing the time gaps inside of it. Same to a first-in-first-out queue on the time axis, this context graph can easily adapt to new trends as time goes by.

Specifically, we build our evolution module by employing two parallel CGNNs \cite{xhonneux2020continuous}, each of which performs historical dynamics evolution or contextual dynamics evolution respectively. The history graph, formed by all happened interactions, captures the relatively static users' preferences and items' attributes. The context graph, formed by all recent interactions inside one context window, captures dynamic time-varying trends and accommodates potential drifts in users' interests and items' status. In this way, the evolution module in CPMR is able to capture evolution dynamics under both history and context scenarios.

To differentiate the message passing capability of different nodes in an interaction graph, we define the learnable spectral radii by a vector $\boldsymbol{\alpha} \in \mathbb{R}^{|\mathcal{U}\cup \mathcal{I}|\times 1}$ for all nodes. Intuitively, each $\alpha_i$ reflects the importance of node $i$ when its embedding is used to form the embeddings of its neighbors. With $\boldsymbol{\alpha}$, we have the learnable adjacency matrix as $ \boldsymbol{A} := \text{Broadcast}\left(\text{Sigmoid}(\boldsymbol{\alpha})\right)\odot \widetilde{\boldsymbol{adj}} $
where Broadcast function expands the vector $\boldsymbol{\alpha}$ to $\mathbb{R}^{|\mathcal{U}\cup \mathcal{I}|\times d}$ by copying, $\odot$ denotes element-wise multiplication, and we employ sigmoid function $\text{Sigmoid}(\boldsymbol{\alpha})$ to scale all eigenvalues of $\boldsymbol{A}$ to be in the interval $[0, 1)$, since $\text{max}\left(\text{Sigmoid}(\boldsymbol{\alpha})\right) < 1$. In subsequent discussions, we refer to the learnable adjacency matrix when the adjacency matrix of a history/context/instant graph is used.

Considering that the same user (item) has different interests (attributes) in different scenarios \cite{wang2020global}, we define the CGNNs' ODEs for the history graph and context graph respectively as follows:
\begin{equation}
\left\{
\begin{aligned}
    \frac{\text{d}\boldsymbol{X}^{his}(t)}{\text{d}t} &= (\boldsymbol{A}^{his}_{t} - \boldsymbol{I})\boldsymbol{X}^{his}(t) + \boldsymbol{E}, \\
    \frac{\text{d}\boldsymbol{X}^{ctx}(t,s)}{\text{d}t} &= (\boldsymbol{A}^{ctx}_{t,s} - \boldsymbol{I})\boldsymbol{X}^{ctx}(t) + \boldsymbol{E},
\end{aligned}
\right.
\label{eq: ode}
\end{equation}
where $\boldsymbol{A}^{his}$ and $\boldsymbol{A}^{ctx}$ use different sets of learnable spectral radii, i.e., $\boldsymbol{\alpha}^{his} \neq \boldsymbol{\alpha}^{ctx}$, to account for the differences between historical and contextual scenarios. With $\boldsymbol{X}^{his}(0)=\boldsymbol{X}^{his}(t_{k-1}^+)$ and $\boldsymbol{X}^{ctx}(0)=\boldsymbol{X}^{ctx}(t_{k-1}^+)$, we have the analytical solutions of the two ODEs for $t>t_{k-1}$ and $\Delta t = t - t_{k-1}$ as follows:
\begin{equation}
\left\{
\begin{aligned}
    &\boldsymbol{X}^{his}(t) =  \\
    &\quad (\boldsymbol{A}^{his}_{t} - \boldsymbol{I})^{-1} \left( e^{(\boldsymbol{A}^{his}_{t} - \boldsymbol{I})\Delta t} - \boldsymbol{I}\right) \boldsymbol{E} + e^{(\boldsymbol{A}^{his}_{t} - \boldsymbol{I})\Delta t} \cdot \boldsymbol{X}^{his}(t_{k-1}^+), \\
    &\boldsymbol{X}^{ctx}(t) = \\
    &\quad (\boldsymbol{A}^{ctx}_{t,s} - \boldsymbol{I})^{-1} \left( e^{(\boldsymbol{A}^{ctx}_{t,s} - \boldsymbol{I})\Delta t} - \boldsymbol{I}\right) \boldsymbol{E} + e^{(\boldsymbol{A}^{ctx}_{t,s} - \boldsymbol{I})\Delta t} \cdot \boldsymbol{X}^{ctx}(t_{k-1}^+), \\
\end{aligned}
\right.
\label{eq: ode_sol}
\end{equation}
where we fix $\boldsymbol{A}^{ctx}_{t, s}$ during the entire interval for computation, i.e., $\boldsymbol{A}^{ctx}_{t, s} =\boldsymbol{A}^{ctx}_{t_k^-, s}$ for $t \in (t_{k-1}, t_k)$. This is considered reasonable as the length of the intervals between adjacent batches of interactions is small compared to the length of the context window. With Eq. (\ref{eq: ode_sol}), we can directly apply approximations of matrix inverse \cite{xhonneux2020continuous} and matrix exponential \cite{zhang2021cope} to obtain a discrete solution.

\subsection{Fusion Module}
As shown in Figure \ref{fig:model}.B, we design the fusion module on top of the Customized Gate Control (CGC) model which is the one-layer version of the well-known Progressive Layered Extraction (PLE) model \cite{tang2020progressive}. In CPMR, we create two fusion module instantiations for users and items separately. Taking the instantiation for users as an example, at $t=t_{k}$, the inputs are users' historical temporal states $\boldsymbol{X}_U^{his}(t_{k}^-)$ and contextual temporal states  $\boldsymbol{X}_U^{ctx}(t_{k}^-)$, and the outputs are their updated historical temporal states $\boldsymbol{X}_U^{his}(t_{k})$, updated contextual temporal states  $\boldsymbol{X}_U^{ctx}(t_{k})$, and final representations $\boldsymbol{Z}_U(t_{k})$ for recommendation. 

To be more specific, we implement one shared expert network, three task-specific expert networks, and three gating networks in Figure \ref{fig:model}.B. Each gating network takes the outputs of its corresponding task-specific expert network and the shared expert network to generate task-specific output. By this design, the shared expert network will be affected by all tasks during optimization, but the task-specific expert networks will only be affected by their own tasks. To selectively combine information from shared and task-specific experts, each gating network learns the weights of each expert and sums the experts' output up with these weights. In the instantiation for users, the generation of users' historical temporal states can be summarized as follows:
\begin{equation}
\begin{gathered}
\boldsymbol{X}^{in}_U = \boldsymbol{X}^{his}_U(t_k^-)\ \oconcat\ \boldsymbol{X}^{ctx}_U(t_k^-), \\
w^{his}_U = \text{Softmax}\left (\text{Gate}^{his}_U(\boldsymbol{X}^{in}_U) \right), \\
\boldsymbol{X}^{his}_U(t_k) =  w^{his}_U \cdot \left [ \text{Expert}^{his}_U(\boldsymbol{X}^{in}_U), \ \text{Expert}^{shared}_U(\boldsymbol{X}^{in}_U) \right ]^T,\\
\end{gathered}
\end{equation}
where $\oconcat$ denotes tensor concatenation on the latent dimension, $\boldsymbol{X}^{in}_U$ denotes concatenated tensor input, $w^{his}_U$ denotes the weights of shared and task-specific expert networks learned from the gating network $\text{Gate}^{his}_U$, $\text{Expert}^{shared}_U$ denotes the shared expert network, $\text{Expert}^{his}_U$ and $\text{Gate}^{his}_U$ denote the expert and gating networks for the task of generating the updated users' historical temporal states. All output terms, including final representations, and historical and contextual temporal states for both users and items, can be generated by the two fusion module instantiations respectively in a similar way.

\subsection{Update Module}
As evidenced in previous SR models \cite{wu2017recurrent, kang2018self}, the appearance of a new interaction has an instant impact on its corresponding user and item within a short period of time. Compared to continuous evolution, this process can be seen as discrete. The update module in our CPMR is designed to implement this sudden change in users' and items' states. At each timestamp $t=t_k$ when some interactions are made, the update module takes the set of new concurrent interactions $\mathcal{E}^{ins}_{t_k}$ as input and transforms the temporal states from $\boldsymbol{X}(t_k)$ to $\boldsymbol{X}(t^+_k)$, which is performed in a discrete way to reflect the instant impact. As shown in Figure \ref{fig:model}.C, given the bi-adjacency matrix of the instant graph $\mathcal{G}^{ins}_{t_k}$, $\boldsymbol{B}^{ins}_{t_k}$, the contextual temporal states update procedure for users is formulated as:
\begin{equation}
\begin{aligned}
    \Delta \boldsymbol{X}^{ctx}_U(t_k) =&\ \text{ReLU}\left((\boldsymbol{D}^{ins}_{t_k, U})^{-1} \cdot \boldsymbol{W}^{ctx}_{i2u} \cdot \boldsymbol{B}^{ins}_{t_k}\cdot \boldsymbol{X}^{ctx}_{I}(t_k)\right), \\
    \boldsymbol{X}^{ctx}_{U}(t_k^+) =&\ \boldsymbol{W}^{ctx}_{u2u} \cdot \boldsymbol{X}^{ctx}_{U}(t_k) + \boldsymbol{M}_{t_k, U} \odot \Delta \boldsymbol{X}^{ctx}_{U}(t_k),
\end{aligned}
\label{eq:update_u}
\end{equation}
where $\boldsymbol{W}^{ctx}_{i2u}$ and $\boldsymbol{W}^{ctx}_{u2u}$ are learnable weight matrices, $\boldsymbol{D}^{ins}_{t_k, U} \in \mathbb{R}^{|\mathcal{U}|\times|\mathcal{U}|}$ is the diagonal degree matrix on rows of $\boldsymbol{B}^{ins}_{t_k}$, and $\boldsymbol{M}_{t_k, U}\in \{0,1\}^{|\mathcal{U}|\times d}$ is a masking matrix that gives entries of 1 to the users involved in $\mathcal{G}^{ins}_{t_k}$. Other temporal states of users and items can be updated in a similar way.

\subsection{Predict Module and Optimization}
To provide a top-N recommendation list for each user who makes interactions at $t=t_k$, we calculate the inner-product similarity $\lambda$ of each user to all items. Given a user $u$ (an item $i$) at $t=t_k$, we fuse its static embedding $\boldsymbol{E}_u$ ($\boldsymbol{E}_i$) and its final representation $\boldsymbol{Z}_u(t_k)$ ($\boldsymbol{Z}_i(t_k)$) with one linear layer, and compute user-item similarity by inner product:
\begin{equation}
\begin{aligned}
\lambda(u, i, t_k) &= \text{FC}_U\left(\boldsymbol{E}_u\ \oconcat\ \boldsymbol{Z}_{u}(t_k)\right) \cdot \text{FC}_I\left(\boldsymbol{E}_i\ \oconcat\ \boldsymbol{Z}_i (t_k)\right),
\end{aligned}
\end{equation}
where $\oconcat$ denotes concatenation on the latent dimension, $\text{FC}_U$ and $\text{FC}_I$ are two fully-connected layers.

\noindent
\textbf{Optimization} During the model optimization, TBPTT is performed every 20 batches (i.e., 20 unique timestamps). For each interaction $(u, i, t_k)$ in $\mathcal{E}^{ins}_{t_k}$, we randomly sample $N_{neg}$ negative users that have never interacted with the item $i$ before $t_k$, and $N_{neg}$ negative items that user $u$ has never interacted with before $t_k$. We then create a negative edge set $\mathcal{E}^{neg}_{t_k}$ by connecting sampled negative users to $i$, and sampled negative items to $u$. We compute the prediction loss for each interaction $(u, i, t_k) \in \mathcal{E}^{ins}_{t_k}$ via InfoNCE \cite{DBLP:journals/corr/abs-1807-03748}:
\begin{equation}
\mathcal{L}(u, i, t_k) = - \text{log}\left(\frac{e^{\lambda(u, i, t_k)}}{e^{\lambda(u, i, t_k)} + \sum_{(u', i', t_k)\in \mathcal{E}^{neg}_{t_k}} e^{\lambda(u', i',t_k)}}\right).
\label{eq:loss}
\end{equation}
The incremental recommendation loss at $t=t_k$ is computed as the average loss over all concurrent interactions in $\mathcal{G}^{ins}_{t_k}$:
\begin{equation}
\mathcal{L}_{t_k} = \frac{1}{\left|\mathcal{E}^{ins}_{t_k}\right|}\sum_{(u, i, t_k)\in \mathcal{E}^{ins}_{t_k}}\mathcal{L}(u, i, t_k).
\label{eq:step_loss}
\end{equation}

\section{Experiments}
In this section, we conduct experiments to evaluate the effectiveness of CPMR. We aim to answer the following questions.
\begin{itemize}
    \item{\textbf{RQ1}:} How good is the performance of CPMR when compared with state-of-the-art evolution models?
    \item{\textbf{RQ2}:} How does the length of the context window affect the performance of CPMR?
    \item{\textbf{RQ3}:} How does the proposed PMTL paradigm affect joint optimization and the mutual update of temporal states?
    \item{\textbf{RQ4}:} How do the proposed contextual temporal states affect the performance of CPMR compared to historical temporal states?
\end{itemize}

\begin{table}[t]
  \caption{Statistics of Sequential Recommendation Datasets.}
  \label{tab:data}
  \begin{tabular}{ccccc}
    \toprule
    & Garden & Video & Games & ML-100K \\
    \midrule
    \# Users & 1,686 & 5,130 & 24,303 & 943 \\
    \# Items & 962 & 1,685 & 10,672 & 1,349 \\
    \# Interactions & 13,272 & 37,126 & 231,780 & 99,287 \\
    \# Timestamps & 1,888 & 1,946 & 5,302 & 49,119 \\
    Span (Days) & 5,221 & 4,984 & 5,395 & 214.83 \\
    \bottomrule
  \end{tabular}
\end{table}
\begin{table}[h]
\caption{Hyperparameters in proposed CPMR.}
\centering
\begin{tabular}{c|c}
\toprule
Hyperparameter & Value \\
\midrule
Embedding dimension $d$ & 128 \\
\# negative users / items $N_{neg}$ & 8 \\
Max epoch & 50 \\
Optimizer & Adam \\
\# batches per TBPTT $n_{tbptt}$ & 20 \\
\midrule
Learning rate & \{5., 2., 1.\} $\times$ \{1e-2, 1e-3, 1e-4, 1e-5\} \\
Weight decay & \{5., 2., 1.\} $\times$ \{1e-2, 1e-3, 1e-4, 1e-5\} \\
Learning rate decay & $\times$ \{0.5, 0.2\} every \{6, 10\} epochs \\
Context window length $s$ & Search from 5 to 100 at a step of 5 \\
\bottomrule
\end{tabular}
\label{tab:hyperparameter}
\end{table}
\begin{table*}[t]
  \caption{Recommendation performance. R@10 is short for Recall@10. The results of all baselines in the first section are imported from CoPE \cite{zhang2021cope}. The results of CoPE (ours) in the second section are derived from our rewritten code. The best results and the runner-up among baselines and CPMR are highlighted in \textbf{bold} and \underline{underline} respectively. The \% Gains are calculated by comparing the best-performing baseline with CPMR. Statistical significance of pairwise differences of CMPR vs. the best baseline is determined by a paired t-test ($***$, $**$ for p-value $\leq 0.01, 0.05$ respectively).}
  \begin{tabular}{c|cc|cc|cc|cc|cc}
    \toprule
     & \multicolumn{2}{c}{Garden} & \multicolumn{2}{c}{Video} & \multicolumn{2}{c}{Game} & \multicolumn{2}{c}{ML-100K} \\
     & MRR & R@10 & MRR & R@10 & MRR & R@10 & MRR & R@10 \\
    \midrule
    LightGCN & 0.025 & 0.087 & 0.019 & 0.036 & 0.015 & 0.026 & 0.012 & 0.025 \\
    Time-LSTM & 0.038 & 0.134 & 0.028 & 0.044 & 0.014 & 0.020 & 0.022 & 0.058 \\
    RRN & 0.072 & 0.152 & 0.033 & 0.068 & 0.018 & 0.029 & 0.032 & 0.065 \\
    DeepCoevolve & 0.046 & 0.121 & 0.023 & 0.050 & 0.013 & 0.027 & 0.029 & 0.069 \\
    JODIE* & 0.049 & 0.127 & 0.037 & 0.078 & 0.021 & 0.035 & 0.034 & 0.074 \\
    CoPE* & 0.081 & 0.192 & \underline{0.048} & 0.088 & 0.026 & 0.047 & 0.038 & 0.081 \\
    \midrule
    CoPE (ours) & \underline{0.0844}\tiny{± 0.0012} & \underline{0.1953}\tiny{± 0.0029} & 0.0421\tiny{± 0.0010} & \underline{0.0922}\tiny{± 0.0014} & \underline{0.0302}\tiny{± 0.0005} & \underline{0.0585}\tiny{± 0.0013} & \underline{0.0443}\tiny{± 0.0008} & \underline{0.0954}\tiny{± 0.0020} \\
    \midrule
    CPMR & \textbf{0.0853}\tiny{± 0.0009} & \textbf{0.2021}\tiny{± 0.0034} & \textbf{0.0664}\tiny{± 0.0019} & \textbf{0.1327}\tiny{± 0.0022} & \textbf{0.0445}\tiny{± 0.0022} & \textbf{0.0842}\tiny{± 0.0034} & \textbf{0.0522}\tiny{± 0.0023} & \textbf{0.1128}\tiny{± 0.0073} \\
    \midrule
    \% Gain & 1.06\% & 3.48\%$^{**}$ & 57.71\%$^{***}$ & 43.92\%$^{***}$ & 47.35\%$^{***}$ & 43.93\%$^{***}$ & 17.83\%$^{***}$ & 18.23\%$^{***}$ \\
    \bottomrule
  \end{tabular}
\label{tab:rq1}
\end{table*}

\begin{figure*}[t]
\centering
\setlength{\abovecaptionskip}{0.1cm}
\includegraphics[width=\linewidth]{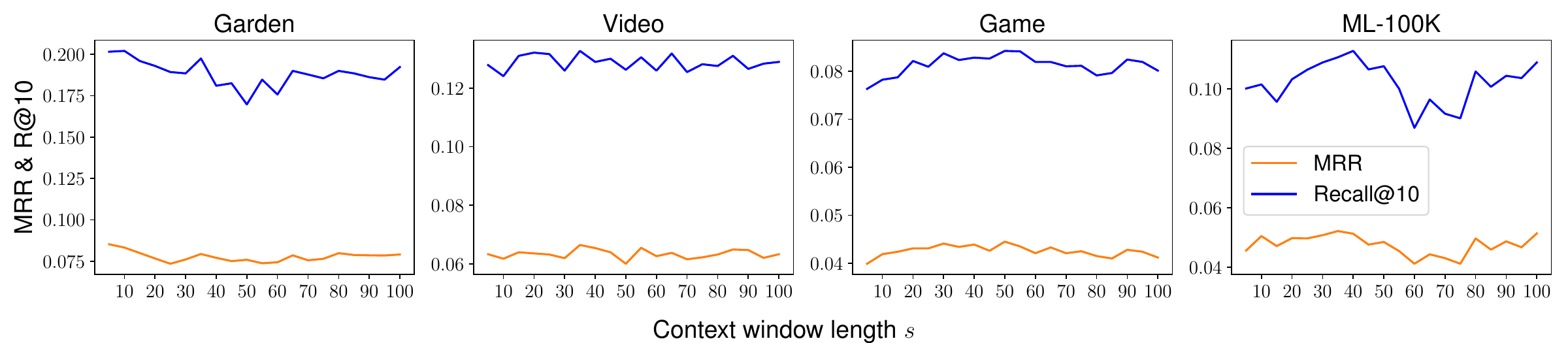}
\caption{Results on Garden, Video, Game and ML-100K w.r.t. different context window length $s$ (in days).}
\label{fig:len_ctx}
\end{figure*}

\subsection{Experimental Setup}
\textbf{Datasets:} We conduct experiments on four public sequential recommendation datasets, including three subsets from Amazon review datasets \cite{he2016ups} (`Patio, Lawn and Garden' as Garden, `Amazon Instant Video' as Video, and `Video Games' as Games)\footnote{\url{http://jmcauley.ucsd.edu/data/amazon/index.html}}, and one movie rating dataset (ML-100K)\footnote{\url{https://grouplens.org/datasets/movielens/}}. The reason for not using session-based recommendation datasets is that long-term sequences of user interactions can show clearer interest evolution compared to short-term anonymous sessions, and session-based recommendation models usually do not explicitly embed users. We use the same preprocessing pipeline of Caser \cite{tang2018personalized} and CoPE \cite{zhang2021cope}, which recursively discards users and items with less than 5 observations until each remaining user or item contains at least 5 interactions. By following CoPE \cite{zhang2021cope} and JODIE \cite{kumar2019predicting}, each dataset is split by time into 80\%/10\%/10\% as training, validation and test sets. The statistics of these datasets are summarized in Table \ref{tab:data}. During experiments, we coarse the timestamp into day to make a trade-off between the granularity and efficiency of the incremental recommendation. We run CPMR and CoPE (ours) five times with different seeds on each dataset to obtain the experimental results.

\textbf{Baselines:} 
Regarding the baseline selection and quality metrics used, we follow those in CoPE \cite{zhang2021cope}. Specifically, we compare CPMR with various baselines, including (1) graph-based recommendation model, LightGCN \cite{he2020lightgcn}; (2) deep recurrent recommendation models, such as Time-LSTM \cite{zhu2017next} and RRN \cite{wu2017recurrent}; (3) temporal network embedding models, such as DeepCoevolve \cite{dai2016deep}, JODIE \cite{kumar2019predicting} and CoPE \cite{zhang2021cope}. For JODIE and CoPE, we use their variants reported in the paper of CoPE \cite{zhang2021cope}: JODIE* that disables test-time training and CoPE* that disables test-time training, meta-learning, and jump loss. We also report our own rewritten and re-tuned CoPE (ours) with meta-learning and jump loss implemented. For quality metrics, we use the MRR (mean reciprocal rank) for the targets among all items, and Recall@10 scores for target items among top-10 recommendations. For a fair comparison, we use the reported results from CoPE \cite{zhang2021cope} for all baselines in the first section in Table \ref{tab:rq1}. 

\noindent\textbf{Hyperparameters:} Our selections of hyperparameters are reported in Table \ref{tab:hyperparameter}. We use $d=128$ as the dimensions of static embeddings and two temporal states for CPMR and CoPE (ours). Considering the fact that incremental recommendation is very prone to overfitting, we carefully tune these models on learning rate and l-2 regularization weight (weight decay) using small step sizes. To figure out how context affects model performance, we also tune the window length from 5 days to 100 days at a step of 5 days.

\subsection{Recommendation Performance (\textbf{RQ1})}
As shown in Table \ref{tab:rq1}, CPMR outperforms all baselines on both MRR and Recall@10 across all four sequential recommendation datasets for ISR tasks. Compared with the best-performing baseline models, CPMR achieves 30.98\% gains on MRR and 27.39\% gains on Recall@10 on average. Statistics significance from paired t-value tests show that CPMR outperforms CoPE significantly on Video, Game and ML-100K datasets where the contextual trends are much more obvious. For Garden, we will explain why the gain is subtle in subsequent sections. The superior performance of CPMR over CoPE is credited to the design of the PMTL paradigm and context awareness. Other temporal baselines perform poorer than CoPE and CPMR because of their recurrent treatment on concurrent interactions. Without time awareness, LightGCN lacks the capability of capturing embedding dynamics, resulting in its inferiority in comparison.

\subsection{Length of Context Window (\textbf{RQ2})}  \label{sec: len_ctx}

The length of the context window $s$ is an important hyperparameter affecting the context awareness of CPMR. Figure \ref{fig:len_ctx} reports the results of this sensitivity study on four datasets. We use context window lengths $s$ in multiples of 5, ranging from 5 to 100. The optimal length will be discussed below.

\noindent\textbf{Garden:} The optimal $s$ is 5 days, but the total time span is 5221 days. Such a short context length suggests that hardly any contextual trends exist. Considering that garden tools iterate very slowly, a shorter $s$ matches the reality. The lack of context also results in the smallest improvement ratio on Garden across all datasets. 

\noindent\textbf{ML-100K \& Video:} The optimal $s$ is 35 days. Since both datasets contain movies and TV series, the length is roughly in line with their showtime and popularity cycle on social networks as well. 

\noindent\textbf{Game:} The optimal $s$ is 50 days. Typically, gaming communities can engage in prolonged discussions about popular games, while meticulously crafted games often demand tens to hundreds of hours for completion. In this case, it makes sense to use a longer contextual window.

\begin{table}[ht]
  \caption{Ablation studies of the fusion module and contextual/historical temporal states.}
  \begin{tabular}{c|cc|cc}
    \toprule
     & \multicolumn{2}{c}{Garden} & \multicolumn{2}{c}{Video} \\
     & MRR & R@10 & MRR & R@10 \\
    \midrule
    CPMR & \textbf{0.0853}\tiny{± 0.0009} & \textbf{0.2021}\tiny{± 0.0034} & \textbf{0.0664}\tiny{± 0.0019} & \textbf{0.1327}\tiny{± 0.0022} \\
    \midrule
    w/o ctx & 0.0789\tiny{± 0.0020} & 0.1935\tiny{± 0.0060} & 0.0565\tiny{± 0.0025} & 0.1078\tiny{± 0.0035} \\
    w/o his & 0.0804\tiny{± 0.0017} & 0.1950\tiny{± 0.0026} & 0.0613\tiny{± 0.0013} & 0.1262\tiny{± 0.0015} \\
    \midrule
    w/o fusion & 0.0838\tiny{± 0.0048} & 0.2021\tiny{± 0.0045} & 0.0611\tiny{± 0.0008} & 0.1171\tiny{± 0.0024} \\
    \bottomrule
  \end{tabular}
\label{tab:ablation}
\end{table}

\subsection{Ablation: Context and History (RQ3)}\label{sec: ablation_context}
To validate the effectiveness of contextual awareness, we choose the
two datasets with the largest and smallest CPMR improvements,
Garden and Video, and design two variants for the ablation studies as
follows:
\begin{itemize}
    \item \textbf{w/o ctx}: disable contextual temporal states by removing the contextual instantiation of the evolution module and changing the input of the fusion module into historical temporal states only (including removing tensor concatenation, $\text{Gate}^{ctx}$ network and $\text{Expert}^{ctx}$ network).
    \item \textbf{w/o his}: disable historical temporal states by removing the historical instantiation of the evolution module and changing the input of the fusion module into contextual temporal states only (including removing tensor concatenation, $\text{Gate}^{his}$ network and $\text{Expert}^{his}$ network).
\end{itemize}
From the second section of Table \ref{tab:ablation}, removing either of the two temporal states leads to a decrease in model performance. This is because one single temporal state contains less information on the dynamics and cannot conduct information fusion in the fusion module. Furthermore, for both datasets, CPMR without history performs better than the version without context, which shows the importance of context awareness when modeling evolution dynamics.

\subsection{Ablation: Fusion Module (RQ4)} \label{sec: ablation_fusion}
\begin{figure}[t]
\centering
\setlength{\abovecaptionskip}{0.1cm}
\includegraphics[width=\linewidth]{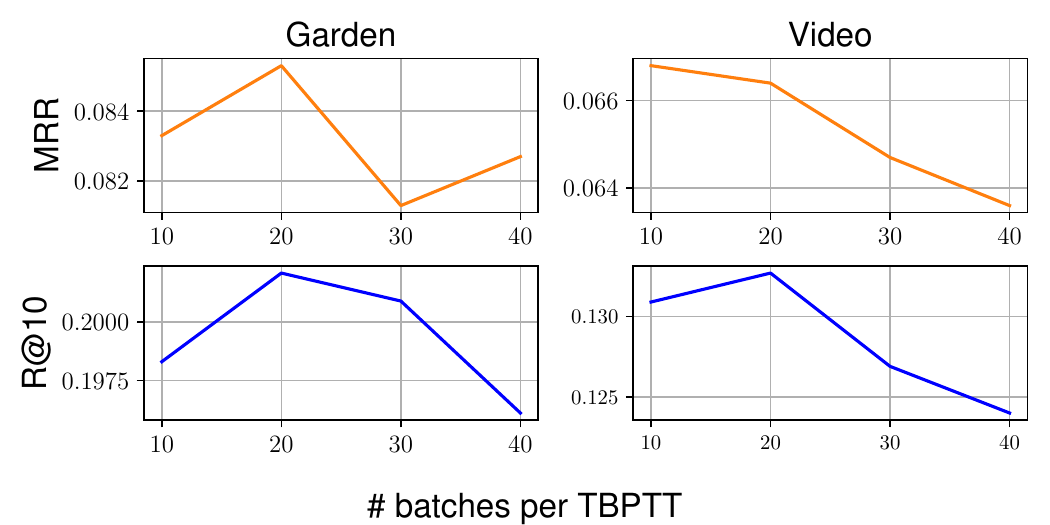}
\caption{Ablation study on \# batches per TBPTT.}
\label{fig:n_tbptt}
\vspace{-0.2cm}
\end{figure}
To validate the effectiveness of the fusion module, we design a variant for this ablation study as follows:
\begin{itemize}
    \item \textbf{w/o fusion}: remove the fusion module, directly adding involved temporal states $\boldsymbol{Z}(t_k) = \boldsymbol{X}(t_k) = \boldsymbol{X}^{his}(t_k^-) + \boldsymbol{X}^{ctx}(t_k^-)$ as input of update module and predict module.
\end{itemize}
We can observe from Table \ref{tab:ablation} that, without the fusion module, the performance of CPMR drops dramatically on Video, but only slightly on Garden. Based on the experiment in Section \ref{sec: len_ctx}, we believe this is because the context of the garden contains too little unique information, which makes the information fusion less effective.

To better inspect the performance of the fusion module, we also tune the number of batches per TBPTT, $n_{tbptt}$, as they can be seen as the number of tasks in MTL. The results in Figure \ref{fig:n_tbptt} show that choosing an appropriate number is important, as fewer batches may lead to inefficiency and task-specific overfitting, while more batches may also lead to a lack of guidance on temporal state evolution. We observe that setting $n_{tbptt}=20$ is a good choice for both Garden and Video datasets.

\section{Conclusion}
In this paper, we propose a novel recommender system, CPMR, that is equipped with both context-aware temporal dynamics modeling and Pseudo-Multi-Task Learning. By jointly optimizing multi-target incremental recommendations, CPMR is able to effectively capture and fuse historical and contextual temporal dynamic states. With such designs, our approach outperforms state-of-the-art models by 30.98\% on MRR and 27.39\% on Recall@10 on average, as demonstrated by extensive experimental studies.

Since the introduction of context shows great potential, in the future, we plan to enrich context with other information (e.g., social links with timestamps, cross-domain sequences with timestamps).

 \bibliographystyle{ACM-Reference-Format}
\bibliography{citation}


\begin{thebibliography}{52}


\ifx \showCODEN    \undefined \def \showCODEN     #1{\unskip}     \fi
\ifx \showDOI      \undefined \def \showDOI       #1{#1}\fi
\ifx \showISBNx    \undefined \def \showISBNx     #1{\unskip}     \fi
\ifx \showISBNxiii \undefined \def \showISBNxiii  #1{\unskip}     \fi
\ifx \showISSN     \undefined \def \showISSN      #1{\unskip}     \fi
\ifx \showLCCN     \undefined \def \showLCCN      #1{\unskip}     \fi
\ifx \shownote     \undefined \def \shownote      #1{#1}          \fi
\ifx \showarticletitle \undefined \def \showarticletitle #1{#1}   \fi
\ifx \showURL      \undefined \def \showURL       {\relax}        \fi
\providecommand\bibfield[2]{#2}
\providecommand\bibinfo[2]{#2}
\providecommand\natexlab[1]{#1}
\providecommand\showeprint[2][]{arXiv:#2}

\bibitem[Bai et~al\mbox{.}(2022)]%
        {bai2022contrastive}
\bibfield{author}{\bibinfo{person}{Ting Bai}, \bibinfo{person}{Yudong Xiao},
  \bibinfo{person}{Bin Wu}, \bibinfo{person}{Guojun Yang},
  \bibinfo{person}{Hongyong Yu}, {and} \bibinfo{person}{Jian-Yun Nie}.}
  \bibinfo{year}{2022}\natexlab{}.
\newblock \showarticletitle{A Contrastive Sharing Model for Multi-Task
  Recommendation}. In \bibinfo{booktitle}{\emph{Proceedings of the ACM Web
  Conference 2022}}. \bibinfo{pages}{3239--3247}.
\newblock


\bibitem[Bai et~al\mbox{.}(2019)]%
        {bai2019ctrec}
\bibfield{author}{\bibinfo{person}{Ting Bai}, \bibinfo{person}{Lixin Zou},
  \bibinfo{person}{Wayne~Xin Zhao}, \bibinfo{person}{Pan Du},
  \bibinfo{person}{Weidong Liu}, \bibinfo{person}{Jian-Yun Nie}, {and}
  \bibinfo{person}{Ji-Rong Wen}.} \bibinfo{year}{2019}\natexlab{}.
\newblock \showarticletitle{CTrec: A long-short demands evolution model for
  continuous-time recommendation}. In \bibinfo{booktitle}{\emph{Proceedings of
  the 42nd International ACM SIGIR Conference on Research and Development in
  Information Retrieval}}. \bibinfo{pages}{675--684}.
\newblock


\bibitem[Brost et~al\mbox{.}(2019)]%
        {brost2019music}
\bibfield{author}{\bibinfo{person}{Brian Brost}, \bibinfo{person}{Rishabh
  Mehrotra}, {and} \bibinfo{person}{Tristan Jehan}.}
  \bibinfo{year}{2019}\natexlab{}.
\newblock \showarticletitle{The music streaming sessions dataset}. In
  \bibinfo{booktitle}{\emph{The World Wide Web Conference}}.
  \bibinfo{pages}{2594--2600}.
\newblock


\bibitem[Chang et~al\mbox{.}(2021)]%
        {chang2021sequential}
\bibfield{author}{\bibinfo{person}{Jianxin Chang}, \bibinfo{person}{Chen Gao},
  \bibinfo{person}{Yu Zheng}, \bibinfo{person}{Yiqun Hui},
  \bibinfo{person}{Yanan Niu}, \bibinfo{person}{Yang Song},
  \bibinfo{person}{Depeng Jin}, {and} \bibinfo{person}{Yong Li}.}
  \bibinfo{year}{2021}\natexlab{}.
\newblock \showarticletitle{Sequential recommendation with graph neural
  networks}. In \bibinfo{booktitle}{\emph{Proceedings of the 44th International
  ACM SIGIR Conference on Research and Development in Information Retrieval}}.
  \bibinfo{pages}{378--387}.
\newblock


\bibitem[Cho et~al\mbox{.}(2020)]%
        {cho2020meantime}
\bibfield{author}{\bibinfo{person}{Sung~Min Cho}, \bibinfo{person}{Eunhyeok
  Park}, {and} \bibinfo{person}{Sungjoo Yoo}.} \bibinfo{year}{2020}\natexlab{}.
\newblock \showarticletitle{MEANTIME: Mixture of attention mechanisms with
  multi-temporal embeddings for sequential recommendation}. In
  \bibinfo{booktitle}{\emph{Fourteenth ACM Conference on Recommender Systems}}.
  \bibinfo{pages}{515--520}.
\newblock


\bibitem[Chung(1997)]%
        {chung1997spectral}
\bibfield{author}{\bibinfo{person}{Fan~RK Chung}.}
  \bibinfo{year}{1997}\natexlab{}.
\newblock \bibinfo{booktitle}{\emph{Spectral graph theory}}.
  Vol.~\bibinfo{volume}{92}.
\newblock \bibinfo{publisher}{American Mathematical Soc.}
\newblock


\bibitem[Dai et~al\mbox{.}(2016)]%
        {dai2016deep}
\bibfield{author}{\bibinfo{person}{Hanjun Dai}, \bibinfo{person}{Yichen Wang},
  \bibinfo{person}{Rakshit Trivedi}, {and} \bibinfo{person}{Le Song}.}
  \bibinfo{year}{2016}\natexlab{}.
\newblock \showarticletitle{Deep coevolutionary network: Embedding user and
  item features for recommendation}.
\newblock \bibinfo{journal}{\emph{arXiv preprint arXiv:1609.03675}}
  (\bibinfo{year}{2016}).
\newblock


\bibitem[De~Domenico et~al\mbox{.}(2013)]%
        {de2013anatomy}
\bibfield{author}{\bibinfo{person}{Manlio De~Domenico},
  \bibinfo{person}{Antonio Lima}, \bibinfo{person}{Paul Mougel}, {and}
  \bibinfo{person}{Mirco Musolesi}.} \bibinfo{year}{2013}\natexlab{}.
\newblock \showarticletitle{The anatomy of a scientific rumor}.
\newblock \bibinfo{journal}{\emph{Scientific reports}} \bibinfo{volume}{3},
  \bibinfo{number}{1} (\bibinfo{year}{2013}), \bibinfo{pages}{1--9}.
\newblock


\bibitem[Donkers et~al\mbox{.}(2017)]%
        {donkers2017sequential}
\bibfield{author}{\bibinfo{person}{Tim Donkers}, \bibinfo{person}{Benedikt
  Loepp}, {and} \bibinfo{person}{J{\"u}rgen Ziegler}.}
  \bibinfo{year}{2017}\natexlab{}.
\newblock \showarticletitle{Sequential user-based recurrent neural network
  recommendations}. In \bibinfo{booktitle}{\emph{Proceedings of the eleventh
  ACM conference on recommender systems}}. \bibinfo{pages}{152--160}.
\newblock


\bibitem[Fan et~al\mbox{.}(2021)]%
        {fan2021continuous}
\bibfield{author}{\bibinfo{person}{Ziwei Fan}, \bibinfo{person}{Zhiwei Liu},
  \bibinfo{person}{Jiawei Zhang}, \bibinfo{person}{Yun Xiong},
  \bibinfo{person}{Lei Zheng}, {and} \bibinfo{person}{Philip~S Yu}.}
  \bibinfo{year}{2021}\natexlab{}.
\newblock \showarticletitle{Continuous-time sequential recommendation with
  temporal graph collaborative transformer}. In
  \bibinfo{booktitle}{\emph{Proceedings of the 30th ACM International
  Conference on Information \& Knowledge Management}}.
  \bibinfo{pages}{433--442}.
\newblock


\bibitem[Gao et~al\mbox{.}(2019)]%
        {gao2019neural}
\bibfield{author}{\bibinfo{person}{Chen Gao}, \bibinfo{person}{Xiangnan He},
  \bibinfo{person}{Dahua Gan}, \bibinfo{person}{Xiangning Chen},
  \bibinfo{person}{Fuli Feng}, \bibinfo{person}{Yong Li},
  \bibinfo{person}{Tat-Seng Chua}, {and} \bibinfo{person}{Depeng Jin}.}
  \bibinfo{year}{2019}\natexlab{}.
\newblock \showarticletitle{Neural multi-task recommendation from
  multi-behavior data}. In \bibinfo{booktitle}{\emph{2019 IEEE 35th
  international conference on data engineering (ICDE)}}. IEEE,
  \bibinfo{pages}{1554--1557}.
\newblock


\bibitem[He and McAuley(2016)]%
        {he2016ups}
\bibfield{author}{\bibinfo{person}{Ruining He} {and} \bibinfo{person}{Julian
  McAuley}.} \bibinfo{year}{2016}\natexlab{}.
\newblock \showarticletitle{Ups and downs: Modeling the visual evolution of
  fashion trends with one-class collaborative filtering}. In
  \bibinfo{booktitle}{\emph{proceedings of the 25th international conference on
  world wide web}}. \bibinfo{pages}{507--517}.
\newblock


\bibitem[He et~al\mbox{.}(2020)]%
        {he2020lightgcn}
\bibfield{author}{\bibinfo{person}{Xiangnan He}, \bibinfo{person}{Kuan Deng},
  \bibinfo{person}{Xiang Wang}, \bibinfo{person}{Yan Li},
  \bibinfo{person}{Yongdong Zhang}, {and} \bibinfo{person}{Meng Wang}.}
  \bibinfo{year}{2020}\natexlab{}.
\newblock \showarticletitle{Lightgcn: Simplifying and powering graph
  convolution network for recommendation}. In
  \bibinfo{booktitle}{\emph{Proceedings of the 43rd International ACM SIGIR
  conference on research and development in Information Retrieval}}.
  \bibinfo{pages}{639--648}.
\newblock


\bibitem[Kang and McAuley(2018)]%
        {kang2018self}
\bibfield{author}{\bibinfo{person}{Wang-Cheng Kang} {and}
  \bibinfo{person}{Julian McAuley}.} \bibinfo{year}{2018}\natexlab{}.
\newblock \showarticletitle{Self-attentive sequential recommendation}. In
  \bibinfo{booktitle}{\emph{2018 IEEE international conference on data mining
  (ICDM)}}. IEEE, \bibinfo{pages}{197--206}.
\newblock


\bibitem[Kumar et~al\mbox{.}(2019)]%
        {kumar2019predicting}
\bibfield{author}{\bibinfo{person}{Srijan Kumar}, \bibinfo{person}{Xikun
  Zhang}, {and} \bibinfo{person}{Jure Leskovec}.}
  \bibinfo{year}{2019}\natexlab{}.
\newblock \showarticletitle{Predicting dynamic embedding trajectory in temporal
  interaction networks}. In \bibinfo{booktitle}{\emph{Proceedings of the 25th
  ACM SIGKDD international conference on knowledge discovery \& data mining}}.
  \bibinfo{pages}{1269--1278}.
\newblock


\bibitem[Li et~al\mbox{.}(2020)]%
        {li2020time}
\bibfield{author}{\bibinfo{person}{Jiacheng Li}, \bibinfo{person}{Yujie Wang},
  {and} \bibinfo{person}{Julian McAuley}.} \bibinfo{year}{2020}\natexlab{}.
\newblock \showarticletitle{Time interval aware self-attention for sequential
  recommendation}. In \bibinfo{booktitle}{\emph{Proceedings of the 13th
  international conference on web search and data mining}}.
  \bibinfo{pages}{322--330}.
\newblock


\bibitem[Ma et~al\mbox{.}(2020)]%
        {ma2020memory}
\bibfield{author}{\bibinfo{person}{Chen Ma}, \bibinfo{person}{Liheng Ma},
  \bibinfo{person}{Yingxue Zhang}, \bibinfo{person}{Jianing Sun},
  \bibinfo{person}{Xue Liu}, {and} \bibinfo{person}{Mark Coates}.}
  \bibinfo{year}{2020}\natexlab{}.
\newblock \showarticletitle{Memory augmented graph neural networks for
  sequential recommendation}. In \bibinfo{booktitle}{\emph{Proceedings of the
  AAAI conference on artificial intelligence}}, Vol.~\bibinfo{volume}{34}.
  \bibinfo{pages}{5045--5052}.
\newblock


\bibitem[Ma et~al\mbox{.}(2018b)]%
        {ma2018modeling}
\bibfield{author}{\bibinfo{person}{Jiaqi Ma}, \bibinfo{person}{Zhe Zhao},
  \bibinfo{person}{Xinyang Yi}, \bibinfo{person}{Jilin Chen},
  \bibinfo{person}{Lichan Hong}, {and} \bibinfo{person}{Ed~H Chi}.}
  \bibinfo{year}{2018}\natexlab{b}.
\newblock \showarticletitle{Modeling task relationships in multi-task learning
  with multi-gate mixture-of-experts}. In \bibinfo{booktitle}{\emph{Proceedings
  of the 24th ACM SIGKDD international conference on knowledge discovery \&
  data mining}}. \bibinfo{pages}{1930--1939}.
\newblock


\bibitem[Ma et~al\mbox{.}(2018a)]%
        {ma2018entire}
\bibfield{author}{\bibinfo{person}{Xiao Ma}, \bibinfo{person}{Liqin Zhao},
  \bibinfo{person}{Guan Huang}, \bibinfo{person}{Zhi Wang},
  \bibinfo{person}{Zelin Hu}, \bibinfo{person}{Xiaoqiang Zhu}, {and}
  \bibinfo{person}{Kun Gai}.} \bibinfo{year}{2018}\natexlab{a}.
\newblock \showarticletitle{Entire space multi-task model: An effective
  approach for estimating post-click conversion rate}. In
  \bibinfo{booktitle}{\emph{The 41st International ACM SIGIR Conference on
  Research \& Development in Information Retrieval}}.
  \bibinfo{pages}{1137--1140}.
\newblock


\bibitem[Rendle et~al\mbox{.}(2010)]%
        {rendle2010factorizing}
\bibfield{author}{\bibinfo{person}{Steffen Rendle}, \bibinfo{person}{Christoph
  Freudenthaler}, {and} \bibinfo{person}{Lars Schmidt-Thieme}.}
  \bibinfo{year}{2010}\natexlab{}.
\newblock \showarticletitle{Factorizing personalized markov chains for
  next-basket recommendation}. In \bibinfo{booktitle}{\emph{Proceedings of the
  19th international conference on World wide web}}. \bibinfo{pages}{811--820}.
\newblock


\bibitem[Song et~al\mbox{.}(2019)]%
        {song2019session}
\bibfield{author}{\bibinfo{person}{Weiping Song}, \bibinfo{person}{Zhiping
  Xiao}, \bibinfo{person}{Yifan Wang}, \bibinfo{person}{Laurent Charlin},
  \bibinfo{person}{Ming Zhang}, {and} \bibinfo{person}{Jian Tang}.}
  \bibinfo{year}{2019}\natexlab{}.
\newblock \showarticletitle{Session-based social recommendation via dynamic
  graph attention networks}. In \bibinfo{booktitle}{\emph{Proceedings of the
  Twelfth ACM international conference on web search and data mining}}.
  \bibinfo{pages}{555--563}.
\newblock


\bibitem[Sun et~al\mbox{.}(2019)]%
        {sun2019bert4rec}
\bibfield{author}{\bibinfo{person}{Fei Sun}, \bibinfo{person}{Jun Liu},
  \bibinfo{person}{Jian Wu}, \bibinfo{person}{Changhua Pei},
  \bibinfo{person}{Xiao Lin}, \bibinfo{person}{Wenwu Ou}, {and}
  \bibinfo{person}{Peng Jiang}.} \bibinfo{year}{2019}\natexlab{}.
\newblock \showarticletitle{BERT4Rec: Sequential recommendation with
  bidirectional encoder representations from transformer}. In
  \bibinfo{booktitle}{\emph{Proceedings of the 28th ACM international
  conference on information and knowledge management}}.
  \bibinfo{pages}{1441--1450}.
\newblock


\bibitem[Sun et~al\mbox{.}(2020)]%
        {sun2020learning}
\bibfield{author}{\bibinfo{person}{Tianxiang Sun}, \bibinfo{person}{Yunfan
  Shao}, \bibinfo{person}{Xiaonan Li}, \bibinfo{person}{Pengfei Liu},
  \bibinfo{person}{Hang Yan}, \bibinfo{person}{Xipeng Qiu}, {and}
  \bibinfo{person}{Xuanjing Huang}.} \bibinfo{year}{2020}\natexlab{}.
\newblock \showarticletitle{Learning sparse sharing architectures for multiple
  tasks}. In \bibinfo{booktitle}{\emph{Proceedings of the AAAI conference on
  artificial intelligence}}, Vol.~\bibinfo{volume}{34}.
  \bibinfo{pages}{8936--8943}.
\newblock


\bibitem[Tang et~al\mbox{.}(2020)]%
        {tang2020progressive}
\bibfield{author}{\bibinfo{person}{Hongyan Tang}, \bibinfo{person}{Junning
  Liu}, \bibinfo{person}{Ming Zhao}, {and} \bibinfo{person}{Xudong Gong}.}
  \bibinfo{year}{2020}\natexlab{}.
\newblock \showarticletitle{Progressive layered extraction (ple): A novel
  multi-task learning (mtl) model for personalized recommendations}. In
  \bibinfo{booktitle}{\emph{Proceedings of the 14th ACM Conference on
  Recommender Systems}}. \bibinfo{pages}{269--278}.
\newblock


\bibitem[Tang and Wang(2018)]%
        {tang2018personalized}
\bibfield{author}{\bibinfo{person}{Jiaxi Tang} {and} \bibinfo{person}{Ke
  Wang}.} \bibinfo{year}{2018}\natexlab{}.
\newblock \showarticletitle{Personalized top-n sequential recommendation via
  convolutional sequence embedding}. In \bibinfo{booktitle}{\emph{Proceedings
  of the eleventh ACM international conference on web search and data mining}}.
  \bibinfo{pages}{565--573}.
\newblock


\bibitem[van~den Oord et~al\mbox{.}(2018)]%
        {DBLP:journals/corr/abs-1807-03748}
\bibfield{author}{\bibinfo{person}{A{\"{a}}ron van~den Oord},
  \bibinfo{person}{Yazhe Li}, {and} \bibinfo{person}{Oriol Vinyals}.}
  \bibinfo{year}{2018}\natexlab{}.
\newblock \showarticletitle{Representation Learning with Contrastive Predictive
  Coding}.
\newblock \bibinfo{journal}{\emph{CoRR}}  \bibinfo{volume}{abs/1807.03748}
  (\bibinfo{year}{2018}).
\newblock
\showeprint[arXiv]{1807.03748}
\urldef\tempurl%
\url{http://arxiv.org/abs/1807.03748}
\showURL{%
\tempurl}


\bibitem[Vass{\o}y et~al\mbox{.}(2019)]%
        {vassoy2019time}
\bibfield{author}{\bibinfo{person}{Bj{\o}rnar Vass{\o}y},
  \bibinfo{person}{Massimiliano Ruocco}, \bibinfo{person}{Eliezer de~Souza~da
  Silva}, {and} \bibinfo{person}{Erlend Aune}.}
  \bibinfo{year}{2019}\natexlab{}.
\newblock \showarticletitle{Time is of the essence: a joint hierarchical rnn
  and point process model for time and item predictions}. In
  \bibinfo{booktitle}{\emph{Proceedings of the Twelfth ACM International
  Conference on Web Search and Data Mining}}. \bibinfo{pages}{591--599}.
\newblock


\bibitem[Wang et~al\mbox{.}(2021)]%
        {wang2021sequential}
\bibfield{author}{\bibinfo{person}{Dongjing Wang}, \bibinfo{person}{Xin Zhang},
  \bibinfo{person}{Zhengzhe Xiang}, \bibinfo{person}{Dongjin Yu},
  \bibinfo{person}{Guandong Xu}, {and} \bibinfo{person}{Shuiguang Deng}.}
  \bibinfo{year}{2021}\natexlab{}.
\newblock \showarticletitle{Sequential Recommendation Based on Multivariate
  Hawkes Process Embedding With Attention}.
\newblock \bibinfo{journal}{\emph{IEEE transactions on cybernetics}}
  (\bibinfo{year}{2021}).
\newblock


\bibitem[Wang et~al\mbox{.}(2019)]%
        {wang2019multi}
\bibfield{author}{\bibinfo{person}{Hongwei Wang}, \bibinfo{person}{Fuzheng
  Zhang}, \bibinfo{person}{Miao Zhao}, \bibinfo{person}{Wenjie Li},
  \bibinfo{person}{Xing Xie}, {and} \bibinfo{person}{Minyi Guo}.}
  \bibinfo{year}{2019}\natexlab{}.
\newblock \showarticletitle{Multi-task feature learning for knowledge graph
  enhanced recommendation}. In \bibinfo{booktitle}{\emph{The world wide web
  conference}}. \bibinfo{pages}{2000--2010}.
\newblock


\bibitem[Wang et~al\mbox{.}(2020a)]%
        {wang2020next}
\bibfield{author}{\bibinfo{person}{Jianling Wang}, \bibinfo{person}{Kaize
  Ding}, \bibinfo{person}{Liangjie Hong}, \bibinfo{person}{Huan Liu}, {and}
  \bibinfo{person}{James Caverlee}.} \bibinfo{year}{2020}\natexlab{a}.
\newblock \showarticletitle{Next-item recommendation with sequential
  hypergraphs}. In \bibinfo{booktitle}{\emph{Proceedings of the 43rd
  international ACM SIGIR conference on research and development in information
  retrieval}}. \bibinfo{pages}{1101--1110}.
\newblock


\bibitem[Wang et~al\mbox{.}(2022)]%
        {wang2022rete}
\bibfield{author}{\bibinfo{person}{Ruijie Wang}, \bibinfo{person}{Zheng Li},
  \bibinfo{person}{Danqing Zhang}, \bibinfo{person}{Qingyu Yin},
  \bibinfo{person}{Tong Zhao}, \bibinfo{person}{Bing Yin}, {and}
  \bibinfo{person}{Tarek Abdelzaher}.} \bibinfo{year}{2022}\natexlab{}.
\newblock \showarticletitle{RETE: Retrieval-Enhanced Temporal Event Forecasting
  on Unified Query Product Evolutionary Graph}. In
  \bibinfo{booktitle}{\emph{Proceedings of the ACM Web Conference 2022}}.
  \bibinfo{pages}{462--472}.
\newblock


\bibitem[Wang et~al\mbox{.}(2020b)]%
        {wang2020practical}
\bibfield{author}{\bibinfo{person}{Yichao Wang}, \bibinfo{person}{Huifeng Guo},
  \bibinfo{person}{Ruiming Tang}, \bibinfo{person}{Zhirong Liu}, {and}
  \bibinfo{person}{Xiuqiang He}.} \bibinfo{year}{2020}\natexlab{b}.
\newblock \showarticletitle{A practical incremental method to train deep ctr
  models}.
\newblock \bibinfo{journal}{\emph{arXiv preprint arXiv:2009.02147}}
  (\bibinfo{year}{2020}).
\newblock


\bibitem[Wang et~al\mbox{.}(2020c)]%
        {wang2020global}
\bibfield{author}{\bibinfo{person}{Ziyang Wang}, \bibinfo{person}{Wei Wei},
  \bibinfo{person}{Gao Cong}, \bibinfo{person}{Xiao-Li Li},
  \bibinfo{person}{Xian-Ling Mao}, {and} \bibinfo{person}{Minghui Qiu}.}
  \bibinfo{year}{2020}\natexlab{c}.
\newblock \showarticletitle{Global context enhanced graph neural networks for
  session-based recommendation}. In \bibinfo{booktitle}{\emph{Proceedings of
  the 43rd international ACM SIGIR conference on research and development in
  information retrieval}}. \bibinfo{pages}{169--178}.
\newblock


\bibitem[Wu et~al\mbox{.}(2017)]%
        {wu2017recurrent}
\bibfield{author}{\bibinfo{person}{Chao-Yuan Wu}, \bibinfo{person}{Amr Ahmed},
  \bibinfo{person}{Alex Beutel}, \bibinfo{person}{Alexander~J Smola}, {and}
  \bibinfo{person}{How Jing}.} \bibinfo{year}{2017}\natexlab{}.
\newblock \showarticletitle{Recurrent recommender networks}. In
  \bibinfo{booktitle}{\emph{Proceedings of the tenth ACM international
  conference on web search and data mining}}. \bibinfo{pages}{495--503}.
\newblock


\bibitem[Wu et~al\mbox{.}(2019a)]%
        {wu2019feature}
\bibfield{author}{\bibinfo{person}{Qitian Wu}, \bibinfo{person}{Lei Jiang},
  \bibinfo{person}{Xiaofeng Gao}, \bibinfo{person}{Xiaochun Yang}, {and}
  \bibinfo{person}{Guihai Chen}.} \bibinfo{year}{2019}\natexlab{a}.
\newblock \showarticletitle{Feature Evolution Based Multi-Task Learning for
  Collaborative Filtering with Social Trust.}. In
  \bibinfo{booktitle}{\emph{IJCAI}}. \bibinfo{pages}{3877--3883}.
\newblock


\bibitem[Wu et~al\mbox{.}(2019b)]%
        {wu2019session}
\bibfield{author}{\bibinfo{person}{Shu Wu}, \bibinfo{person}{Yuyuan Tang},
  \bibinfo{person}{Yanqiao Zhu}, \bibinfo{person}{Liang Wang},
  \bibinfo{person}{Xing Xie}, {and} \bibinfo{person}{Tieniu Tan}.}
  \bibinfo{year}{2019}\natexlab{b}.
\newblock \showarticletitle{Session-based recommendation with graph neural
  networks}. In \bibinfo{booktitle}{\emph{Proceedings of the AAAI conference on
  artificial intelligence}}, Vol.~\bibinfo{volume}{33}.
  \bibinfo{pages}{346--353}.
\newblock


\bibitem[Xhonneux et~al\mbox{.}(2020)]%
        {xhonneux2020continuous}
\bibfield{author}{\bibinfo{person}{Louis-Pascal Xhonneux},
  \bibinfo{person}{Meng Qu}, {and} \bibinfo{person}{Jian Tang}.}
  \bibinfo{year}{2020}\natexlab{}.
\newblock \showarticletitle{Continuous graph neural networks}. In
  \bibinfo{booktitle}{\emph{International Conference on Machine Learning}}.
  PMLR, \bibinfo{pages}{10432--10441}.
\newblock


\bibitem[Xia et~al\mbox{.}(2022)]%
        {xia2022fire}
\bibfield{author}{\bibinfo{person}{Jiafeng Xia}, \bibinfo{person}{Dongsheng
  Li}, \bibinfo{person}{Hansu Gu}, \bibinfo{person}{Jiahao Liu},
  \bibinfo{person}{Tun Lu}, {and} \bibinfo{person}{Ning Gu}.}
  \bibinfo{year}{2022}\natexlab{}.
\newblock \showarticletitle{FIRE: Fast Incremental Recommendation with Graph
  Signal Processing}. In \bibinfo{booktitle}{\emph{Proceedings of the ACM Web
  Conference 2022}}. \bibinfo{pages}{2360--2369}.
\newblock


\bibitem[Xu et~al\mbox{.}(2019b)]%
        {xu2019recurrent}
\bibfield{author}{\bibinfo{person}{Chengfeng Xu}, \bibinfo{person}{Pengpeng
  Zhao}, \bibinfo{person}{Yanchi Liu}, \bibinfo{person}{Jiajie Xu},
  \bibinfo{person}{Victor S~Sheng S.~Sheng}, \bibinfo{person}{Zhiming Cui},
  \bibinfo{person}{Xiaofang Zhou}, {and} \bibinfo{person}{Hui Xiong}.}
  \bibinfo{year}{2019}\natexlab{b}.
\newblock \showarticletitle{Recurrent convolutional neural network for
  sequential recommendation}. In \bibinfo{booktitle}{\emph{The world wide web
  conference}}. \bibinfo{pages}{3398--3404}.
\newblock


\bibitem[Xu et~al\mbox{.}(2019a)]%
        {xu2019self}
\bibfield{author}{\bibinfo{person}{Da Xu}, \bibinfo{person}{Chuanwei Ruan},
  \bibinfo{person}{Evren Korpeoglu}, \bibinfo{person}{Sushant Kumar}, {and}
  \bibinfo{person}{Kannan Achan}.} \bibinfo{year}{2019}\natexlab{a}.
\newblock \showarticletitle{Self-attention with functional time representation
  learning}.
\newblock \bibinfo{journal}{\emph{Advances in neural information processing
  systems}}  \bibinfo{volume}{32} (\bibinfo{year}{2019}).
\newblock


\bibitem[Yang et~al\mbox{.}(2022)]%
        {yang2022stam}
\bibfield{author}{\bibinfo{person}{Zhen Yang}, \bibinfo{person}{Ming Ding},
  \bibinfo{person}{Bin Xu}, \bibinfo{person}{Hongxia Yang}, {and}
  \bibinfo{person}{Jie Tang}.} \bibinfo{year}{2022}\natexlab{}.
\newblock \showarticletitle{STAM: A Spatiotemporal Aggregation Method for Graph
  Neural Network-based Recommendation}. In
  \bibinfo{booktitle}{\emph{Proceedings of the ACM Web Conference 2022}}.
  \bibinfo{pages}{3217--3228}.
\newblock


\bibitem[Ying et~al\mbox{.}(2018)]%
        {ying2018sequential}
\bibfield{author}{\bibinfo{person}{Haochao Ying}, \bibinfo{person}{Fuzhen
  Zhuang}, \bibinfo{person}{Fuzheng Zhang}, \bibinfo{person}{Yanchi Liu},
  \bibinfo{person}{Guandong Xu}, \bibinfo{person}{Xing Xie},
  \bibinfo{person}{Hui Xiong}, {and} \bibinfo{person}{Jian Wu}.}
  \bibinfo{year}{2018}\natexlab{}.
\newblock \showarticletitle{Sequential recommender system based on hierarchical
  attention network}. In \bibinfo{booktitle}{\emph{IJCAI International Joint
  Conference on Artificial Intelligence}}.
\newblock


\bibitem[Zhang et~al\mbox{.}(2019)]%
        {zhang2019feature}
\bibfield{author}{\bibinfo{person}{Tingting Zhang}, \bibinfo{person}{Pengpeng
  Zhao}, \bibinfo{person}{Yanchi Liu}, \bibinfo{person}{Victor~S Sheng},
  \bibinfo{person}{Jiajie Xu}, \bibinfo{person}{Deqing Wang},
  \bibinfo{person}{Guanfeng Liu}, {and} \bibinfo{person}{Xiaofang Zhou}.}
  \bibinfo{year}{2019}\natexlab{}.
\newblock \showarticletitle{Feature-level Deeper Self-Attention Network for
  Sequential Recommendation.}. In \bibinfo{booktitle}{\emph{IJCAI}}.
  \bibinfo{pages}{4320--4326}.
\newblock


\bibitem[Zhang et~al\mbox{.}(2020)]%
        {zhang2020retrain}
\bibfield{author}{\bibinfo{person}{Yang Zhang}, \bibinfo{person}{Fuli Feng},
  \bibinfo{person}{Chenxu Wang}, \bibinfo{person}{Xiangnan He},
  \bibinfo{person}{Meng Wang}, \bibinfo{person}{Yan Li}, {and}
  \bibinfo{person}{Yongdong Zhang}.} \bibinfo{year}{2020}\natexlab{}.
\newblock \showarticletitle{How to retrain recommender system? A sequential
  meta-learning method}. In \bibinfo{booktitle}{\emph{Proceedings of the 43rd
  International ACM SIGIR Conference on Research and Development in Information
  Retrieval}}. \bibinfo{pages}{1479--1488}.
\newblock


\bibitem[Zhang et~al\mbox{.}(2021)]%
        {zhang2021cope}
\bibfield{author}{\bibinfo{person}{Yao Zhang}, \bibinfo{person}{Yun Xiong},
  \bibinfo{person}{Dongsheng Li}, \bibinfo{person}{Caihua Shan},
  \bibinfo{person}{Kan Ren}, {and} \bibinfo{person}{Yangyong Zhu}.}
  \bibinfo{year}{2021}\natexlab{}.
\newblock \showarticletitle{CoPE: Modeling Continuous Propagation and Evolution
  on Interaction Graph}. In \bibinfo{booktitle}{\emph{Proceedings of the 30th
  ACM International Conference on Information \& Knowledge Management}}.
  \bibinfo{pages}{2627--2636}.
\newblock


\bibitem[Zheng et~al\mbox{.}(2022)]%
        {zheng2022disentangling}
\bibfield{author}{\bibinfo{person}{Yu Zheng}, \bibinfo{person}{Chen Gao},
  \bibinfo{person}{Jianxin Chang}, \bibinfo{person}{Yanan Niu},
  \bibinfo{person}{Yang Song}, \bibinfo{person}{Depeng Jin}, {and}
  \bibinfo{person}{Yong Li}.} \bibinfo{year}{2022}\natexlab{}.
\newblock \showarticletitle{Disentangling Long and Short-Term Interests for
  Recommendation}. In \bibinfo{booktitle}{\emph{Proceedings of the ACM Web
  Conference 2022}}. \bibinfo{pages}{2256--2267}.
\newblock


\bibitem[Zhou et~al\mbox{.}(2019)]%
        {zhou2019deep}
\bibfield{author}{\bibinfo{person}{Guorui Zhou}, \bibinfo{person}{Na Mou},
  \bibinfo{person}{Ying Fan}, \bibinfo{person}{Qi Pi}, \bibinfo{person}{Weijie
  Bian}, \bibinfo{person}{Chang Zhou}, \bibinfo{person}{Xiaoqiang Zhu}, {and}
  \bibinfo{person}{Kun Gai}.} \bibinfo{year}{2019}\natexlab{}.
\newblock \showarticletitle{Deep interest evolution network for click-through
  rate prediction}. In \bibinfo{booktitle}{\emph{Proceedings of the AAAI
  conference on artificial intelligence}}, Vol.~\bibinfo{volume}{33}.
  \bibinfo{pages}{5941--5948}.
\newblock


\bibitem[Zhou et~al\mbox{.}(2018)]%
        {zhou2018deep}
\bibfield{author}{\bibinfo{person}{Guorui Zhou}, \bibinfo{person}{Xiaoqiang
  Zhu}, \bibinfo{person}{Chenru Song}, \bibinfo{person}{Ying Fan},
  \bibinfo{person}{Han Zhu}, \bibinfo{person}{Xiao Ma},
  \bibinfo{person}{Yanghui Yan}, \bibinfo{person}{Junqi Jin},
  \bibinfo{person}{Han Li}, {and} \bibinfo{person}{Kun Gai}.}
  \bibinfo{year}{2018}\natexlab{}.
\newblock \showarticletitle{Deep interest network for click-through rate
  prediction}. In \bibinfo{booktitle}{\emph{Proceedings of the 24th ACM SIGKDD
  international conference on knowledge discovery \& data mining}}.
  \bibinfo{pages}{1059--1068}.
\newblock


\bibitem[Zhou et~al\mbox{.}(2021)]%
        {zhou2021temporal}
\bibfield{author}{\bibinfo{person}{Huachi Zhou}, \bibinfo{person}{Qiaoyu Tan},
  \bibinfo{person}{Xiao Huang}, \bibinfo{person}{Kaixiong Zhou}, {and}
  \bibinfo{person}{Xiaoling Wang}.} \bibinfo{year}{2021}\natexlab{}.
\newblock \showarticletitle{Temporal augmented graph neural networks for
  session-based recommendations}. In \bibinfo{booktitle}{\emph{Proceedings of
  the 44th International ACM SIGIR Conference on Research and Development in
  Information Retrieval}}. \bibinfo{pages}{1798--1802}.
\newblock


\bibitem[Zhou et~al\mbox{.}(2020)]%
        {zhou2020improving}
\bibfield{author}{\bibinfo{person}{Kun Zhou}, \bibinfo{person}{Wayne~Xin Zhao},
  \bibinfo{person}{Shuqing Bian}, \bibinfo{person}{Yuanhang Zhou},
  \bibinfo{person}{Ji-Rong Wen}, {and} \bibinfo{person}{Jingsong Yu}.}
  \bibinfo{year}{2020}\natexlab{}.
\newblock \showarticletitle{Improving conversational recommender systems via
  knowledge graph based semantic fusion}. In
  \bibinfo{booktitle}{\emph{Proceedings of the 26th ACM SIGKDD International
  Conference on Knowledge Discovery \& Data Mining}}.
  \bibinfo{pages}{1006--1014}.
\newblock


\bibitem[Zhu et~al\mbox{.}(2019)]%
        {zhu2019dtcdr}
\bibfield{author}{\bibinfo{person}{Feng Zhu}, \bibinfo{person}{Chaochao Chen},
  \bibinfo{person}{Yan Wang}, \bibinfo{person}{Guanfeng Liu}, {and}
  \bibinfo{person}{Xiaolin Zheng}.} \bibinfo{year}{2019}\natexlab{}.
\newblock \showarticletitle{Dtcdr: A framework for dual-target cross-domain
  recommendation}. In \bibinfo{booktitle}{\emph{Proceedings of the 28th ACM
  International Conference on Information and Knowledge Management}}.
  \bibinfo{pages}{1533--1542}.
\newblock


\bibitem[Zhu et~al\mbox{.}(2017)]%
        {zhu2017next}
\bibfield{author}{\bibinfo{person}{Yu Zhu}, \bibinfo{person}{Hao Li},
  \bibinfo{person}{Yikang Liao}, \bibinfo{person}{Beidou Wang},
  \bibinfo{person}{Ziyu Guan}, \bibinfo{person}{Haifeng Liu}, {and}
  \bibinfo{person}{Deng Cai}.} \bibinfo{year}{2017}\natexlab{}.
\newblock \showarticletitle{What to Do Next: Modeling User Behaviors by
  Time-LSTM.}. In \bibinfo{booktitle}{\emph{IJCAI}}, Vol.~\bibinfo{volume}{17}.
  \bibinfo{pages}{3602--3608}.
\newblock


\end{thebibliography}

\end{document}